\documentclass[11pt,onecolumn,draftclsnofoot]{IEEEtran}
\usepackage{graphicx,accents}
\usepackage[cmex10]{amsmath}
\usepackage{amsfonts,amssymb,latexsym,cite,ifthen,multirow,array}
\usepackage[hyphens]{url}
\usepackage{xcolor}
\usepackage[normalem]{ulem}
\usepackage[hang, font=footnotesize,center]{caption}
\newboolean{SEP_FIG_CAPS}\setboolean{SEP_FIG_CAPS}{false}
\newboolean{ONE_COLUMN}\setboolean{ONE_COLUMN}{true}
\ifthenelse{\boolean{SEP_FIG_CAPS}}
{

 \newcommand{\capFrag}[2]{\noindent Fig.~\ref{fig:#1}. #2 \medskip\\}
 \newcommand{\capTable}[2]{\noindent Tab.~\ref{tab:#1}. #2 \medskip\\}
}
{

 \newcommand{\capFrag}[2]{}
 \newcommand{\capTable}[2]{}
}

 \newcommand{\ovec}[1]{\ensuremath{\Bar{\boldsymbol{#1}}}}
 \newcommand{\hvec}[1]{\ensuremath{\Hat{\boldsymbol{#1}}}}
 
 \renewcommand{\vec}[1]{\ensuremath{\boldsymbol{#1}}}

 \DeclareMathOperator{\tr}{tr}

 \DeclareMathOperator{\vect}{vec}

 \DeclareMathOperator*{\argmax}{argmax}

 \renewcommand{\eqref}[1]{(\ref{eq:#1})}
 \newcommand{\Eqref}[1]{Equation~(\ref{eq:#1})}
 
 \newcommand{\figref}[1]{Fig.~\ref{fig:#1}}
 \newcommand{\tabref}[1]{Table~\ref{tab:#1}}
 \newcommand{\secref}[1]{Section~\ref{sec:#1}}
 
 \newcommand{\appref}[1]{Appendix~\ref{app:#1}}

 \newcommand{\textr}[1]{\textcolor{black}{#1}}

 \newcounter{comment}[section]
 
 \newcounter{texthead}[section]

 \newcommand{\fwd}[1]{\accentset{\rightharpoonup}{#1}}
 \newcommand{\bwd}[1]{\accentset{\leftharpoonup}{#1}}
 
 \newcommand{\eps}{\varepsilon}
 \newcommand{\poz}{p_{{}_{10}}}
 \newcommand{\pzo}{p_{{}_{01}}}
 
 \newcommand{\mf}[1]{\mathfrak{#1}}
 \renewcommand{\vect}[1]{\vec{#1}^{(t)}}
 \newcommand{\nt}[1]{{#1}_n^{(t)}}
 \newcommand{\msg}[2]{\nu_{{#1} \to {#2}}}

 \newcounter{threshcounter}
 \renewcommand{\thethreshcounter}{D\arabic{threshcounter}}
 \newcommand{\threshcnt}[1]{\refstepcounter{threshcounter} \label{#1} \thetag{\thethreshcounter}}
 \newcounter{algcounter}
 \renewcommand{\thealgcounter}{A\arabic{algcounter}}
 \newcommand{\algcnt}[1]{\refstepcounter{algcounter} \label{#1} \thetag{\thealgcounter}}
 \newcounter{emcounter}
 \renewcommand{\theemcounter}{E\arabic{emcounter}}
 \newcommand{\emcnt}[1]{\refstepcounter{emcounter} \label{#1} \thetag{\theemcounter}}

 \graphicspath{{./}{}}

\begin{document}
\setlength{\arraycolsep}{0.8mm}

%
\title{Dynamic Compressive Sensing of Time-Varying Signals via Approximate Message Passing}
\author{Justin Ziniel,~\IEEEmembership{Student Member,~IEEE}, and Philip Schniter,~\IEEEmembership{Senior Member,~IEEE}
\thanks{The authors are with the Department of Electrical and Computer Engineering, The Ohio State University, Columbus, Ohio.  E-mail: \{zinielj, schniter\}@ece.osu.edu.}
\thanks{Work supported in part by NSF grant CCF-1018368, DARPA/ONR grant N66001-10-1-4090, and an allocation of computing time from the Ohio Supercomputer Center.}
\thanks{Portions of this work were previously presented at the 2010 Asilomar Conference on Signals, Systems, and Computing \cite{ZPS2010}.}
}

\maketitle

\begin{abstract}
In this work the dynamic compressive sensing (CS) problem of recovering sparse, correlated, time-varying signals from sub-Nyquist, non-adaptive, linear measurements is explored from a Bayesian perspective.  While there has been a handful of previously proposed Bayesian dynamic CS algorithms in the literature, the ability to perform inference on high-dimensional problems in a computationally efficient manner remains elusive.  In response, we propose a probabilistic dynamic CS signal model that captures both amplitude and support correlation structure, and describe an approximate message passing algorithm that performs soft signal estimation and support detection with a computational complexity that is linear in all problem dimensions.  The algorithm, DCS-AMP, can perform either causal filtering or non-causal smoothing, and is capable of learning model parameters adaptively from the data through an expectation-maximization learning procedure.  We provide numerical evidence that DCS-AMP performs within $3$ dB of oracle bounds on synthetic data under a variety of operating conditions.  We further describe the result of applying DCS-AMP to two real dynamic CS datasets, as well as a frequency estimation task, to bolster our claim that DCS-AMP is capable of offering state-of-the-art performance and speed on real-world high-dimensional problems.
\end{abstract}

\section{Introduction}
\label{sec:introduction}
In this work, we consider the \emph{dynamic compressive sensing} (dynamic CS) problem, in which a sparse, vector-valued time series is recovered from a second time series of noisy, sub-Nyquist, linear measurements.  Such a problem finds application in\textr{, e.g.,} dynamic MRI \cite{V2008}, high-speed video capture \cite{ARBV2010}, and underwater channel estimation \cite{LP2007}.

Framed mathematically, the objective of the dynamic CS problem is to recover the time series \linebreak$\{\vec{x}^{(1)}, \ldots, \vec{x}^{(T)}\}$, where $\vect{x} \in \mathbb{C}^N$ is the signal at timestep $t$, from a time series of measurements, $\{\vec{y}^{(1)}, \ldots, \vec{y}^{(T)}\}$.  Each $\vect{y} \in \mathbb{C}^M$ is obtained from the linear measurement process,
\begin{equation}
	\vect{y} = \vect{A} \vect{x} + \vect{e}, \qquad t = 1, \ldots, T,
	\label{eq:measurements}
\end{equation}
with $\vect{e}$ representing corrupting noise.  The measurement matrix $\vect{A}$ (which may be time-varying or time-invariant, i.e., $\vect{A} = \vec{A} \,\, \forall \,\, t$) is known in advance, and is generally wide, leading to an underdetermined system of equations.  The problem is regularized by assuming that $\vect{x}$ is sparse (or compressible),\footnote{Without loss of generality, we assume $\vect{x}$ is sparse/compressible in the canonical basis.  Other sparsifying bases can be incorporated into the measurement matrix $\vect{A}$ without changing our model.} having relatively few non-zero (or large) entries.

In many real-world scenarios, the underlying time-varying sparse signal exhibits substantial temporal correlation.  This temporal correlation may manifest itself in two interrelated ways: \textit{(i)} the support of the signal may change slowly over time \textr{\cite{V2008,AGG2009,ARBV2010,V2010b,DV2010,LV2012}}, and \textit{(ii)} the amplitudes of the large coefficients may vary smoothly in time.

In such scenarios, incorporating an appropriate model of temporal structure into a recovery technique makes it possible to drastically outperform structure-agnostic CS algorithms.  From an analytical standpoint, Vaswani and Lu demonstrate that the restricted isometry property (RIP) sufficient conditions for perfect recovery in the dynamic CS problem are significantly weaker than those found in the traditional single measurement vector (SMV) CS problem when accounting for the additional structure \cite{VL2010}.  In this work, we take a Bayesian approach to modeling this structure, which contrasts those dynamic CS algorithms inspired by convex relaxation, such as the Dynamic LASSO \cite{AGG2009} and the Modified-CS algorithm \cite{VL2010}.  Our Bayesian framework is also distinct from those hybrid techniques that blend elements of Bayesian dynamical models like the Kalman filter with more traditional CS approaches of exploiting sparsity through convex relaxation \cite{V2008, ARG2009} or greedy methods \cite{DSM2011}.

In particular, we propose a probabilistic model that treats the time-varying signal support as a set of independent binary Markov processes and the time-varying coefficient amplitudes as a set of independent Gauss-Markov processes.  As detailed in \secref{signal_model}, this model leads to coefficient marginal distributions that are Bernoulli-Gaussian (i.e., ``spike-and-slab'').  Later, in \secref{structure}, we describe a generalization of the aforementioned model that yields Bernoulli-Gaussian-mixture coefficient marginals with an arbitrary number of mixture components.  The models that we propose thus differ substantially from those used in other Bayesian approaches to dynamic CS, \cite{SAP2010} and \cite{STR2011}.  In particular, Sejdinovi\'c et al. \cite{SAP2010} combine a linear Gaussian dynamical system model with a sparsity-promoting Gaussian-scale-mixture prior, while Shahrasbi et al. \cite{STR2011} employ a particular spike-and-slab Markov model that couples amplitude evolution together with support evolution.

Our inference method also differs from those used in the alternative Bayesian dynamic CS algorithms \cite{SAP2010} and \cite{STR2011}.  In \cite{SAP2010}, Sejdinovi\'c et al. perform inference via a sequential Monte Carlo sampler \cite{RC2004}.  Sequential Monte Carlo techniques are appealing for their applicability to complicated non-linear, non-Gaussian inference tasks like the Bayesian dynamic CS problem.  Nevertheless, there are a number of important practical issues related to selection of the importance distribution, choice of the resampling method, and the number of sample points to track, since in principle one must increase the number of points exponentially over time to combat degeneracy \cite{RC2004}.  Additionally, Monte Carlo techniques can be computationally expensive in high-dimensional inference problems.  An alternative inference procedure that has recently proven successful in a number of applications is loopy belief propagation (LBP) \cite{FM1998}.  In \cite{STR2011}, Shahrasbi et al. extend the conventional LBP method proposed in \cite{BSB2010} for standard CS under a sparse measurement matrix $\vec{A}$ to the case of dynamic CS under sparse $\vect{A}$.  Nevertheless, the confinement to sparse measurement matrices is very restrictive, and, without this restriction, the methods of \cite{BSB2010,STR2011} become computationally intractable.

Our inference procedure is based on the recently proposed framework of approximate message passing (AMP) \cite{DMM2009}, and in particular its ``turbo'' extension \cite{S2010a}.  AMP, an unconventional form of LBP, was originally proposed for standard CS with a dense measurement matrix \cite{DMM2009}, and its noteworthy properties include: \textit{(i)} a rigorous analysis (as $M,N \rightarrow \infty$ with $M/N$ fixed, under i.i.d. \textr{sub-}Gaussian $\vec{A}$) establishing that its solutions are governed by a state-evolution whose fixed points are optimal in several respects \cite{BM2011}, and \textit{(ii)} extremely fast runtimes (as a consequence of the fact that it needs relatively few iterations, each requiring only one multiplication by $\vec{A}$ and its transpose).  The turbo-AMP framework originally proposed in \cite{S2010a} offers a way to extend AMP to structured-sparsity problems such as compressive imaging \cite{SS2012}, joint communication channel/symbol estimation \cite{S2011}, and---as we shall see in 
this work---the dynamic CS problem.

Our work makes several contributions to the existing literature on dynamic CS.  First and foremost, the DCS-AMP algorithm that we develop offers an unrivaled combination of speed (e.g., its computational complexity grows only linearly in the problem dimensions $M$, $N$, and $T$) and reconstruction accuracy, as we demonstrate on both synthetic and real-world signals.  \textr{Ours is the first work to exploit the speed and accuracy of loopy belief propagation (and, in particular, AMP) in the dynamic CS setting, accomplished by embedding AMP within a larger Bayesian inference algorithm.}  Second, we propose an expectation-maximization \cite{DLR1977} procedure to automatically learn the parameters of our statistical model, as described in \secref{em_updates}, avoiding a potentially complicated ``tuning'' problem.  The ability to automatically calibrate algorithm parameters is especially important when working with real-world data, but is not provided by many of the existing dynamic CS algorithms (e.g., \cite{V2008,AGG2009,ARG2009,SAP2010,VL2010,DSM2011}).  In addition, our learned model parameters provide a convenient and interpretable characterization of time-varying signals in a way that, e.g., Lagrange multipliers do not.  Third, DCS-AMP provides a unified means of performing both filtering, where estimates are obtained sequentially using only past observations, and smoothing, where each estimate enjoys the knowledge of past, current, and future observations.  In contrast, the existing dynamic CS schemes can support either filtering, or smoothing, but not both.

\subsection{Notation}
\label{sec:introduction:notation}

Boldfaced lower-case letters, e.g., $\vec{a}$, denote column vectors, while boldfaced upper-case letters, e.g., $\vec{A}$, denote matrices.  The letter $t$ is strictly used to index a timestep, $t=1, 2, \ldots, T$, the letter $n$ is strictly used to index the coefficients of a signal, $n = 1,\ldots,N$, and the letter $m$ is strictly used to index the measurements, $m=1,\ldots,M$.  The superscript ${}^{(t)}$ indicates a timestep-dependent quantity, while a superscript without parentheses, such as ${}^k$, indicates a quantity whose value changes according to some algorithmic iteration index $k$.  Subscript notations such as $\nt{x}$ are used to denote the $n^{th}$ element of the vector $\vect{x}$, while set subscript notation, e.g., $\vect{x}_{\mathcal{S}}$, denotes the sub-vector of $\vect{x}$ consisting of indices contained in $\mathcal{S}$.  The $m^{th}$ row of the matrix $\vec{A}$ is denoted by $\vec{a}_m^\textsf{T}$, an $M$-by-$M$ identity matrix is denoted by $\vec{I}_M$, and a length-$N$ vector of ones is given by $\vec{1}_N$.  Finally, $\mathcal{CN}(\vec{a};\vec{b},\vec{C})$ refers to the circularly symmetric complex normal distribution that is a function of the vector $\vec{a}$, with mean $\vec{b}$ and covariance matrix $\vec{C}$.

\section{Signal Model}
\label{sec:signal_model}

We assume that the measurement process can be accurately described by the linear model of \eqref{measurements}.  We further assume that $\vect{A} \in \mathbb{C}^{M\times N}, \, t = 1,\ldots,T,$ are measurement matrices known in advance, whose columns have been scaled to be of unit norm.\footnote{Our algorithm can be generalized to support $\vect{A}$ without equal-norm columns, \textr{a time-varying number of measurements, $M^{(t)}$}, and real-valued matrices/signals as well.}  We model the noise as a stationary, circularly symmetric, additive white Gaussian noise (AWGN) process, with $\vect{e} \sim \mathcal{CN}(\vec{0}, \sigma_e^2 \vec{I}_M) \,\, \forall \,\, t$.

As noted in \secref{introduction}, the sparse time series, $\{\vect{x}\}_{t=1}^T$, often exhibits a high degree of correlation from one timestep to the next.  In this work, we model this correlation through a slow time-variation of the signal support, and a smooth evolution of the amplitudes of the non-zero coefficients.  To do so, we introduce two hidden random processes, $\{\vect{s}\}_{t=1}^T$ and $\{\vect{\theta}\}_{t=1}^T$.  The binary vector $\vect{s} \in \{0,1\}^N$ describes the support of $\vect{x}$, denoted $\mathcal{S}^{(t)}$, while the vector $\vect{\theta} \in \mathbb{C}^N$ describes the amplitudes of the active elements of $\vect{x}$.  Together, $\vect{s}$ and $\vect{\theta}$ completely characterize $\vect{x}$ as follows:
\begin{equation}
	\nt{x} = \nt{s} \cdot \nt{\theta} \quad \forall n,t.
	\label{eq:x_decomp}
\end{equation}
Therefore, $\nt{s} = 0$ sets $\nt{x} = 0$ and $n \notin \mathcal{S}^{(t)}$, while $\nt{s} = 1$ sets $\nt{x} = \nt{\theta}$ and $n \in \mathcal{S}^{(t)}$.

To model slow changes in the support $\mathcal{S}^{(t)}$ over time, we model the $n^{th}$ coefficient's support across time, $\{\nt{s}\}_{t=1}^T$, as a Markov chain defined by two transition probabilities: $\poz \! \triangleq \! \text{Pr}\{ s_n^{(t)} \!\! = \! 1 | s_n^{(t-1)} \!\! = \! 0 \}$, and $\pzo \triangleq \text{Pr}\{ s_n^{(t)} = 0 | s_n^{(t-1)} = 1 \}$, and employ independent chains across $n = 1,\ldots,N$.  We further assume that each Markov chain operates in steady-state, such that $\text{Pr}\{s_n^{(t)} = 1\} = \lambda \,\, \forall \, n, t$.  This steady-state assumption implies that these Markov chains are completely specified by the parameters $\lambda$ and $\pzo$, which together determine the remaining transition probability $\poz = \lambda \pzo / (1 - \lambda)$.  Depending on how $\pzo$ is chosen, the prior distribution can favor signals that exhibit a nearly static support across time, or it can allow for signal supports that change substantially from timestep to timestep.  For example, it can be shown that $1/\pzo$ specifies the average run length of a sequence of ones in the Markov chains.

The second form of temporal structure that we capture in our signal model is the correlation in active coefficient amplitudes across time.  We model this correlation through independent stationary steady-state Gauss-Markov processes for each $n$, wherein $\{\nt{\theta}\}_{t=1}^T$ evolves in time according to
\begin{equation}
	\theta_n^{(t)} = (1-\alpha) \big(\theta_n^{(t-1)} - \zeta\ \big) + \alpha w_n^{(t)} + \zeta,
	\label{eq:theta_evolve}
\end{equation}
where $\zeta \in \mathbb{C}$ is the mean of the process, $w_n^{(t)} \sim \mathcal{CN}(0, \rho)$ is an i.i.d. circular white Gaussian perturbation, and $\alpha \in [0,1]$ controls the temporal correlation.  At one extreme, $\alpha = 0$, the amplitudes are totally correlated, (i.e., $\theta_n^{(t)} = \theta_n^{(t-1)}$), while at the other extreme, $\alpha = 1$, the amplitudes evolve according to an uncorrelated Gaussian random process with mean $\zeta$.

At this point, we would like to make a few remarks about our signal model.  First, due to \eqref{x_decomp}, it is clear that $p\big(x_n^{(t)} | s_n^{(t)}, \theta_n^{(t)}\big) = \delta\big(x_n^{(t)} - s_n^{(t)} \theta_n^{(t)}\big)$, where $\delta(\cdot)$ is the Dirac delta function.  By marginalizing out $s_n^{(t)}$ and $\theta_n^{(t)}$, one finds that
\begin{equation}
	p(x_n^{(t)}) = (1 - \lambda) \delta(x_n^{(t)}) + \lambda \, \mathcal{CN}(x_n^{(t)}; \zeta, \sigma^2),
	\label{eq:x_prior}
\end{equation}
where $\sigma^2 \triangleq \tfrac{\alpha \rho}{2 - \alpha}$ is the steady-state variance of $\nt{\theta}$.  \Eqref{x_prior} is a Bernoulli-Gaussian or ``spike-and-slab'' distribution, which is an effective sparsity-promoting prior due to the point-mass at $x_n^{(t)} = 0$.  Second, we observe that the amplitude random process, $\{\vec{\theta}^{(t)}\}_{t=1}^T$, evolves independently from the sparsity pattern random process, $\{\vec{s}^{(t)}\}_{t=1}^T$.  As a result of this modeling choice, there can be significant hidden amplitudes $\nt{\theta}$ associated with inactive coefficients (those for which $s_n^{(t)} = 0$).  Consequently, $\nt{\theta}$ should be viewed as the amplitude of $\nt{x}$ \emph{conditioned} on $\nt{s} = 1$.  Lastly, we note that higher-order Markov processes and/or more complex coefficient marginals could be considered within the framework we propose, however, to keep development simple, we restrict our attention to first-order Markov processes and Bernoulli-Gaussian marginals until \secref{structure}, where we describe an extension of the above signal model that yields Bernoulli-Gaussian-mixture marginals.

\section{The DCS-AMP Algorithm}
\label{sec:algorithm}
In this section we will describe the DCS-AMP algorithm, which efficiently and accurately estimates the marginal posterior distributions of $\{\nt{x}\}$, $\{\nt{\theta}\}$, and $\{\nt{s}\}$ from the observed measurements $\{\vec{y}^{(t)}\}_{t=1}^T$, thus enabling both soft estimation and soft support detection.  \textr{The use of soft support information is particularly advantageous, as it means that the algorithm need never make a firm (and possibly erroneous) decision about the support that can propagate errors across many timesteps.}  As mentioned in \secref{introduction}, DCS-AMP can perform either filtering or smoothing.

\begin{figure}
	\begin{center}
		\ifthenelse{\boolean{ONE_COLUMN}}
		{\includegraphics[angle=-90,bb=85 250  390 435,scale=0.70]{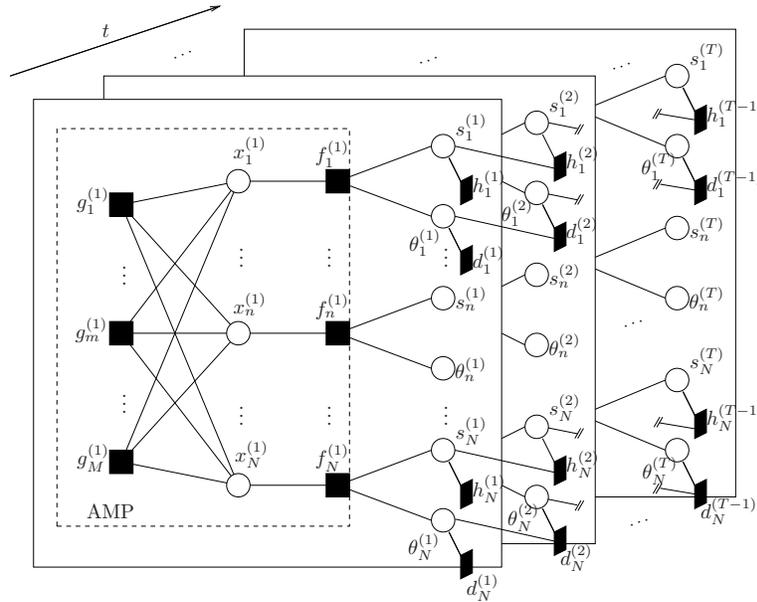}}
		{\includegraphics[angle=-90,bb=85 250  390 435,scale=0.59]{total_factor_graph1}}
		\caption{Factor graph representation of the joint posterior distribution of \eqref{joint_decomp}.}
		\label{fig:total_factor_graph}
	\end{center}
\end{figure}

The algorithm we develop is designed to exploit the statistical structure inherent in our signal model.  By defining $\ovec{y}$ to be the collection of all measurements, $\{\vec{y}^{(t)}\}_{t=1}^T$ (and defining $\ovec{x}$, $\ovec{s}$, and $\ovec{\theta}$ similarly), the posterior joint distribution of the signal, support, and amplitude time series, given the measurement time series, can be expressed using Bayes' rule as
\ifthenelse{\boolean{ONE_COLUMN}}
{
\begin{equation}
	p(\ovec{x}, \ovec{s}, \ovec{\theta} | \ovec{y}) \propto \prod_{t=1}^T \left( \prod_{m=1}^M p(y_m^{(t)} | \vect{x}) \prod_{n=1}^N p(\nt{x} | \nt{s}, \nt{\theta}) p(\nt{s} | s_n^{(t-1)}) p(\nt{\theta} | \theta_n^{(t-1)}) \right),
	\label{eq:joint_decomp}
\end{equation}
}
{
\begin{eqnarray}
	p(\ovec{x}, \ovec{s}, \ovec{\theta} | \ovec{y}) &\propto& \prod_{t=1}^T \Bigg( \prod_{m=1}^M p(y_m^{(t)} | \vect{x}) \prod_{n=1}^N p(\nt{x} | \nt{s}, \nt{\theta})	\nonumber \\
	&\quad& \qquad \times \, p(\nt{s} | s_n^{(t-1)}) p(\nt{\theta} | \theta_n^{(t-1)})\Bigg),
	\label{eq:joint_decomp}
\end{eqnarray}
}
where $\propto$ indicates proportionality up to a constant scale factor, $p(s_n^{(1)} | s_n^{(0)}) \triangleq p(s_n^{(1)})$, and $p(\theta_n^{(1)} | \theta_n^{(0)}) \triangleq p(\theta_n^{(1)})$.  By inspecting \eqref{joint_decomp}, we see that the posterior joint distribution decomposes into the product of many distributions that only depend on small subsets of variables.  A graphical representation of such decompositions is given by the \emph{factor graph}, which is an undirected bipartite graph that connects the pdf ``factors'' of \eqref{joint_decomp} with the random variables that constitute their arguments \cite{KFL2001}.  In \tabref{factors}, we introduce the notation that we will use for the factors of our signal model, showing the correspondence between the factor labels and the underlying distributions they represent, as well as the specific functional form assumed by each factor.  The associated factor graph for the posterior joint distribution of \eqref{joint_decomp} is shown in \figref{total_factor_graph}, labeled according to \tabref{factors}.  Filled squares represent factors, while circles represent random variables.

\begin{table*}
	\begin{center}
	\begin{tabular}{rcl}
		Factor	&	Distribution		&	Functional Form \\
		\hline
		$g_m^{(t)}\big(\vec{x}^{(t)}\big)$	&	$p\big(y_m^{(t)} | \vec{x}^{(t)}\big)$	&	$\mathcal{CN}\big(y_m^{(t)}; \vec{a}_m^{(t) \, T} \vec{x}^{(t)}, \sigma_e^2\big)$	\\
		$f_n^{(t)}\big(x_n^{(t)}, s_n^{(t)}, \theta_n^{(t)}\big)$	&	$p\big(x_n^{(t)} | s_n^{(t)}, \theta_n^{(t)}\big)$	&	$\delta\big(x_n^{(t)} - s_n^{(t)}\theta_n^{(t)}\big)$	\\
		$h_n^{(1)}\big(s_n^{(1)}\big)$	&	$p\big(s_n^{(1)}\big)$	&	$\big(1 - \lambda\big)^{\!1 - s_n^{(1)}\!} \lambda^{s_n^{(1)}}$	\\
		$h_n^{(t)}\big(s_n^{(t)}, s_n^{(t-1)}\big)$	&	$p\big(s_n^{(t)} | s_n^{(t-1)}\big)$	&	
		$\left\{ \begin{array}{ll}
		(1 - \poz)^{\!1 - \nt{s}\!} \poz^{\,\,\,\,\,\, \nt{s}}, & s_n^{(t-1)} = 0 \\
		 \pzo^{\,\,\,\,\,\, 1 - \nt{s}} (1 - \pzo)^{\nt{s}}, & s_n^{(t-1)} = 1
		 \end{array} \right.$\\
		$d_n^{(1)}\big(\theta_n^{(1)}\big)$	&	$p\big(\theta_n^{(1)}\big)$	&	$\mathcal{CN}\big(\theta_n^{(1)}; \zeta, \sigma^2\big)$	\\
		$d_n^{(t)}\big(\theta_n^{(t)}, \theta_n^{(t-1)}\big)$	&	$p\big(\theta_n^{(t)} | \theta_n^{(t-1)}\big)$	&	$\mathcal{CN}\big(\theta_n^{(t)}; (1 - \alpha) \theta_n^{(t-1)} + \alpha \zeta, \alpha^2 \rho \big)$
	\end{tabular}
	\end{center}
	\caption{The factors, underlying distributions, and functional forms associated with our signal model}
	\label{tab:factors}
\end{table*}

As seen in \figref{total_factor_graph}, all of the variables needed at a given timestep can be visualized as lying in a plane, with successive planes stacked one after another in time.  We will refer to these planes as ``frames".  The temporal correlation of the signal supports is illustrated by the $h_n^{(t)}$ factor nodes that connect the $s_n^{(t)}$ variable nodes between neighboring frames.  Likewise, the temporal correlation of the signal amplitudes is expressed by the interconnection of $d_n^{(t)}$ factor nodes and $\theta_n^{(t)}$ variable nodes.  For visual clarity, these factor nodes have been omitted from the middle portion of the factor graph, appearing only at indices $n=1$ and $n=N$, but should in fact be present for all indices $n=1,\ldots,N$.  Since the measurements $\{y_m^{(t)}\}$ are observed variables, they have been incorporated into the $g_m^{(t)}$ factor nodes.

The algorithm that we develop can be viewed as an approximate implementation of belief propagation (BP) \cite{P1988}, a message passing algorithm for performing inference on factor graphs that describe probabilistic models.  When the factor graph is cycle-free, belief propagation is equivalent to the more general sum-product algorithm \cite{KFL2001}, which is a means of computing the marginal functions that result from summing (or integrating) a multivariate function over all possible input arguments, with one argument held fixed, (i.e., marginalizing out all but one variable).  In the context of BP, these marginal functions are the marginal distributions of random variables.  Thus, given measurements $\ovec{y}$ and the factorization of the posterior joint distribution $p\big(\ovec{x}, \ovec{s}, \ovec{\theta} \big| \ovec{y}\big)$, DCS-AMP computes (approximate) posterior marginals of $x_n^{(t)}$, $s_n^{(t)}$, and $\theta_n^{(t)}$.  In ``filtering'' mode, our algorithm would therefore return, e.g., $p\big(x_n^{(t)} \big| \{\vec{y}^{(t)}\}_{t=1}^t\big)$, while in ``smoothing'' mode it would return $p\big(x_n^{(t)} \big| \{\vec{y}^{(t)}\}_{t=1}^T\big)$.  From these marginals, one can compute, e.g., minimum mean-squared error (MMSE) estimates.  The factor graph of \figref{total_factor_graph} contains many short cycles, however, and thus the convergence of loopy BP cannot be guaranteed \cite{KFL2001}.\footnote{However, it is worth noting that in the past decade much work has been accomplished in identifying specific situations under which loopy BP \emph{is} guaranteed to converge, e.g., \cite{WF2001,TJ2002,H2004,IFW2005,BM2011}.}  Despite this, loopy BP has been shown to perform extremely well in a number of different applications, including turbo decoding \cite{MMC1998}, computer vision \cite{FPC2000}, and compressive sensing \cite{DMM2009,BSB2010,S2010a,DMM2010,ZS2013,VS2012,SS2012}.

\subsection{Message scheduling}
\label{sec:algorithm:scheduling}
In loopy factor graphs, there are a number of ways to schedule, or sequence, the messages that are exchanged between nodes.  The choice of a schedule can impact not only the rate of convergence of the algorithm, but also the likelihood of convergence as well \cite{EMK2006}.  We propose a schedule (an evolution of the ``turbo'' schedule proposed in \cite{S2010a}) for DCS-AMP that is straightforward to implement, suitable for both filtering and smoothing applications, and empirically yields quickly converging estimates under a variety of diverse operating conditions.

Our proposed schedule can be broken down into four distinct steps, which we will refer to using the mnemonics \textbf{(into)}, \textbf{(within)}, \textbf{(out)}, and \textbf{(across)}.  At a particular timestep $t$, the \textbf{(into)} step involves passing messages that provide current beliefs about the state of the relevant support variables, $\{\nt{s}\}_{n=1}^N$, and amplitude variables, $\{\nt{\theta}\}_{n=1}^N$, laterally \emph{into} the dashed AMP box within frame $t$.  (Recall \figref{total_factor_graph}.)  The \textbf{(within)} step makes use of these incoming messages, together with the observations available in that frame, $\{y_m^{(t)}\}_{m=1}^M$, to exchange messages \emph{within} the dashed AMP box of frame $t$, thus generating estimates of the marginal posteriors of the signal variables $\{\nt{x}\}_{n=1}^N$.  Using these posterior estimates, the \textbf{(out)} step propagates messages \emph{out} of the dashed AMP box, providing updated beliefs about the state of $\{\nt{s}\}_{n=1}^N$ and $\{\nt{\theta}\}_{n=1}^N$.  Lastly, the \textbf{(across)} step involves transmitting messages \emph{across} neighboring frames, using the updated beliefs about $\{\nt{s}\}_{n=1}^N$ and $\{\nt{\theta}\}_{n=1}^N$ to influence the beliefs about $\{s_n^{(t+1)}\}_{n=1}^N$ and $\{\theta_n^{(t+1)}\}_{n=1}^N$ $\big($or $\{s_n^{(t-1)}\}_{n=1}^N$ and $\{\theta_n^{(t-1)}\}_{n=1}^N$$\big)$.

The procedures for filtering and smoothing both start in the same way.  At the initial $t=1$ frame, steps \textbf{(into)}, \textbf{(within)} and \textbf{(out)} are performed in succession.  Next, step \textbf{(across)} is performed to pass messages from $\{s_n^{(1)}\}_{n=1}^N$ and $\{\theta_n^{(1)}\}_{n=1}^N$ to $\{s_n^{(2)}\}_{n=1}^N$ and $\{\theta_n^{(2)}\}_{n=1}^N$.  Then at frame $t=2$ the same set of steps are executed, concluding with messages propagating to $\{s_n^{(3)}\}_{n=1}^N$ and $\{\theta_n^{(3)}\}_{n=1}^N$.  This process continues until steps \textbf{(into)}, \textbf{(within)} and \textbf{(out)} have been completed at the terminal frame, $T$.  At this point, DCS-AMP has completed what we call a single forward pass.  If the objective was to perform filtering, DCS-AMP terminates at this point, since only causal measurements have been used to estimate the marginal posteriors.  If instead the objective is to obtain smoothed, non-causal estimates, then information begins to propagate backwards in time, i.e., step \textbf{(across)} moves messages from $\{s_n^{(T)}\}_{n=1}^N$ and $\{\theta_n^{(T)}\}_{n=1}^N$ to $\{s_n^{(T-1)}\}_{n=1}^N$ and $\{\theta_n^{(T-1)}\}_{n=1}^N$.  Steps \textbf{(into)}, \textbf{(within)}, \textbf{(out)}, and \textbf{(across)} are performed at frame $T-1$, with messages bound for frame $T-2$.  This continues until the initial frame is reached.  At this point DCS-AMP has completed what we term as a single forward/backward pass.  Multiple such passes, indexed by the variable $k$, can be carried out until a convergence criterion is met or a maximum number of passes has been performed.

\subsection{Implementing the message passes}
\label{sec:algorithm:implementation}

\begin{figure}
	\begin{center}
		\scalebox{0.53}{\rotatebox{90}{\includegraphics*[1.50in,2.75in][6.55in,9.65in]{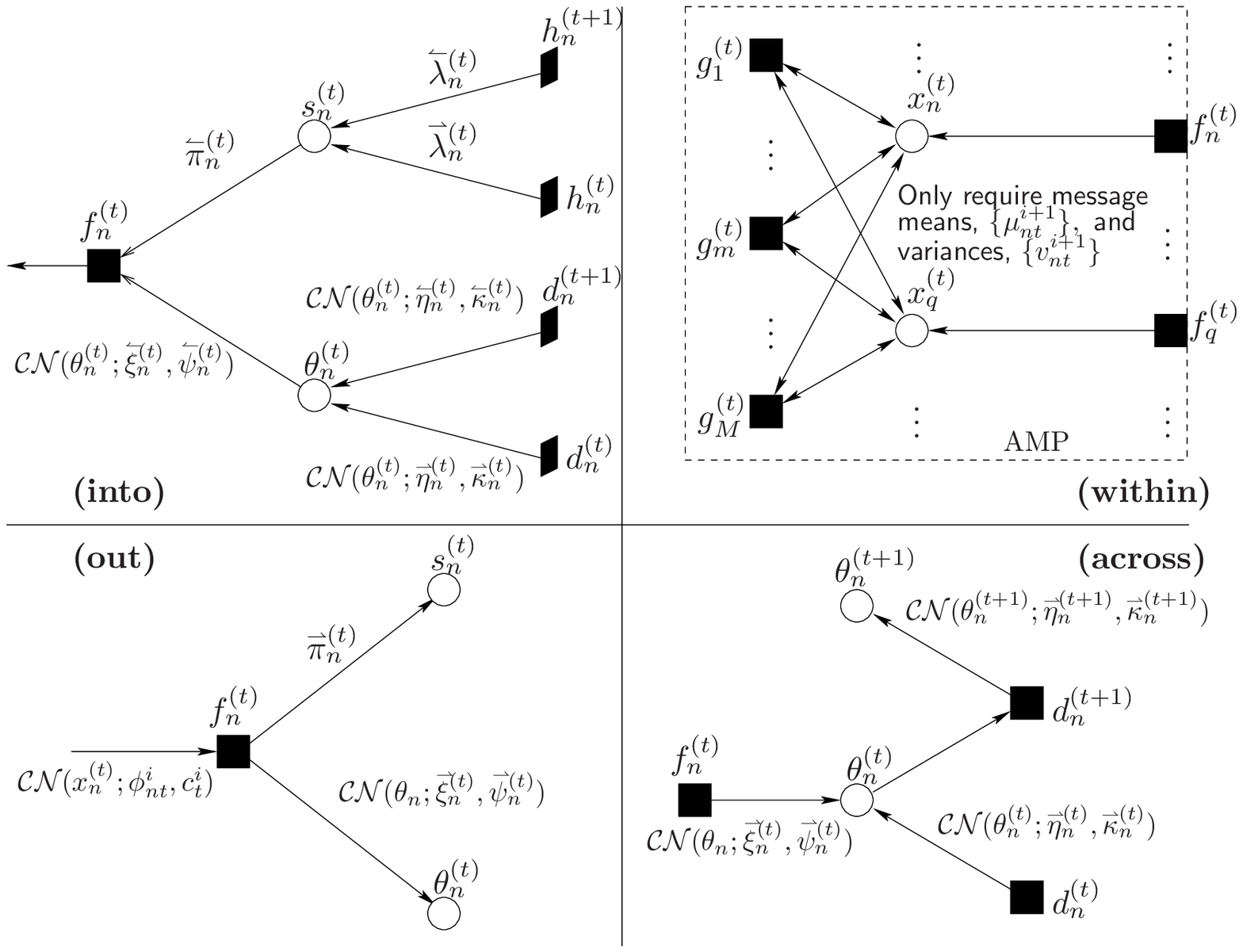}}}
		\caption{A summary of the four message passing phases, including message notation and form.  \textr{See the pseudocode of \tabref{algorithm_equations} for the precise message update computations.}}
		\label{fig:message_summary}
	\end{center}
\end{figure}

We now provide some additional details as to how the above four steps are implemented.  To aid our discussion, in \figref{message_summary} we summarize the form of the messages that pass between the various factor graph nodes, focusing primarily on a single coefficient index $n$ at an intermediate frame $t$.   Directed edges indicate the direction that messages are moving.  In the \textbf{(across)} phase, we only illustrate the messages involved in a forward pass for the amplitude variables, and leave out a graphic for the corresponding backward pass, as well as graphics for the support variable \textbf{(across)} phase.  Note that, to be applicable at frame $T$, the factor node $d_n^{(t+1)}$ and its associated edge should be removed.  The figure also introduces the notation that we adopt for the different variables that serve to parameterize the messages.  \textr{We use the notation $\msg{a}{b}(\cdot)$ to denote a message passing from node $a$ to a connected node $b$.}  For Bernoulli message pdfs, we show only the non-zero probability, e.g., $\nt{\fwd{\lambda}} = \msg{\nt{h}}{\nt{s}}(\nt{s} = 1)$.

To perform step \textbf{(into)}, the messages from the factors $\nt{h}$ and $h_n^{(t+1)}$ to $\nt{s}$ are used to set $\nt{\bwd{\pi}}$, the message from $\nt{s}$ to $\nt{f}$.  Likewise,  the messages from the factors $\nt{d}$ and $d_n^{(t+1)}$ to $\nt{\theta}$ are used to determine the message from $\nt{\theta}$ to $\nt{f}$.  When performing filtering, or the first forward pass of smoothing, no meaningful information should be conveyed from the $h_n^{(t+1)}$ and $d_n^{(t+1)}$ factors.  This can be accomplished by initializing $\big(\nt{\bwd{\lambda}}, \nt{\bwd{\eta}}, \nt{\bwd{\kappa}}\big)$ with the values $(\frac{1}{2}, 0, \infty)$.

In step \textbf{(within)}, messages must be exchanged between the $\big\{\nt{x}\big\}_{n=1}^N$ and $\big\{g_m^{(t)}\big\}_{m=1}^M$ nodes.  When $\vect{A}$ is not a sparse matrix, this will imply a dense network of connections between these nodes.  Consequently, the standard sum-product algorithm would require us to evaluate multi-dimensional integrals of non-Gaussian messages that grow exponentially in number in both $N$ and $M$.  This approach is clearly infeasible for problems of any appreciable size, and thus we turn to a simplification known as \emph{approximate message passing} (AMP) \cite{DMM2009, DMM2010}.

At a high-level, AMP can be viewed as a simplification of loopy BP, employing central limit theorem arguments to approximate the sum of many non-Gaussian random variables as a Gaussian.  Through a series of principled approximation steps (which become exact for \textr{sub-}Gaussian $\vec{A}$ matrices in the large-system limit \cite{BM2011}), AMP produces an iterative thresholding algorithm that requires only $\mathcal{O}(MN)$ operations, dominated by matrix-vector products, to obtain posteriors on the $\big\{\nt{x}\big\}_{n=1}^N$ variable nodes.  The specifics of the iterative thresholding algorithm will depend on the signal prior under which AMP is operating \cite{DMM2010}, but it is assumed that the joint prior decouples into independent (but not necessarily i.i.d.) priors on each coefficient $\nt{x}$.  \textr{See \appref{amp_primer} for additional background on AMP.}

By viewing $\msg{\nt{f}}{\nt{x}}(\cdot)$ as a ``local prior''\footnote{The AMP algorithm is conventionally run with static, i.i.d. priors for each signal coefficient.  When utilized as a sub-component of a larger message passing algorithm on \textr{an expanded} factor graph, the signal priors (from AMP's perspective) will be changing in response to messages from the rest of the factor graph.  We refer to these changing AMP priors as \emph{local priors}.} for $\nt{x}$, we can readily apply \textr{an off-the-shelf} AMP \textr{algorithm (e.g., \cite{DMM2010,R2011,R2010b})} as a means of performing the message passes within the portions of the factor graph enclosed within the dashed boxes of \figref{total_factor_graph} (only one such box is visible).  \textr{The use of AMP with decoupled local priors within a larger message passing algorithm that accounts for statistical dependencies between signal coefficients was first proposed in \cite{S2010a}, and further studied in \cite{S2011,SS2011,SS2012,ZRS2012,ZS2013}.  Here, we exploit this powerful ``turbo'' inference approach to account for the strong temporal dependencies inherent in the dynamic CS problem.}

The local prior \textr{for our signal model} is a Bernoulli-Gaussian, namely
\begin{equation}
	\nu_{f_n^{(t)} \to x_n^{(t)}}(x_n^{(t)}) = (1 - \bwd{\pi}_n^{(t)}) \delta(x_n^{(t)}) + \bwd{\pi}_n^{(t)} \mathcal{CN}(x_n^{(t)}; \bwd{\xi}_n^{(t)}, \bwd{\psi}_n^{(t)}).  \nonumber
\end{equation}
The appropriate AMP message update equations for this local prior follow a straightforward extension of the derivations outlined in \cite{S2010a}, which considered the special case of a zero-mean Bernoulli-Gaussian prior.  The specific AMP updates for our model are given by \eqref{amp_start}-\eqref{amp_end} in \tabref{algorithm_equations}.

After employing AMP to manage the message passing between the $\big\{x_n^{(t)}\big\}_{n=1}^N$ and $\big\{g_m^{(t)}\big\}_{m=1}^M$ nodes in step \textbf{(within)}, messages must be propagated out of the dashed AMP box of frame $t$ (step \textbf{(out)}) and either forward or backward in time (step \textbf{(across)}).  While step \textbf{(across)} simply requires a straightforward application of the sum-product message computation rules, step \textbf{(out)} imposes several difficulties which we must address.   For the remainder of this discussion, we focus on \textr{a novel approximation scheme for specifying} the message $\nu_{f_n^{(t)} \to \theta_n^{(t)}}(\cdot)$.  \textr{Our objective is to arrive at a message approximation that introduces negligible error while still leading to a computationally efficient algorithm.  A Gaussian message approximation is a natural choice, given the marginally Gaussian distribution of $\nt{\theta}$.  As we shall soon see, it is also a highly justifiable choice.}

A routine application of the sum-product rules to the $f_n^{(t)}$\!-to-$\theta_n^{(t)}$ message would produce the following expression:
\begin{equation}
	\msg{\nt{f}}{\nt{\theta}}^{\text{exact}}(\nt{\theta}) \triangleq (1 - \nt{\bwd{\pi}}) \mathcal{CN}(0; \phi_{nt}^i, c_t^i) + \nt{\bwd{\pi}} \mathcal{CN}(\nt{\theta}; \phi_{nt}^i, c_t^i).
	\label{eq:f_to_theta_msg_exact}
\end{equation}
Unfortunately, the term $\mathcal{CN}(0; \phi_{nt}^i, c_t^i)$ prevents us from normalizing $\nu_{f_n^{(t)} \to \theta_n^{(t)}}^{\text{exact}}(\nt{\theta})$, \textr{because it} is constant with respect to $\nt{\theta}$.  Therefore, the distribution on $\theta_n^{(t)}$ represented by \eqref{f_to_theta_msg_exact} is improper.  To provide intuition into why this is the case, it is helpful to think of $\nu_{f_n^{(t)} \to \theta_n^{(t)}}(\nt{\theta})$ as a message that conveys information about the value of $\theta_n^{(t)}$ based on the values of $x_n^{(t)}$ and $s_n^{(t)}$.  If $s_n^{(t)} = 0$, then by \eqref{x_decomp}, $x_n^{(t)} = 0$, thus making $\nt{\theta}$ unobservable.  The constant term in \eqref{f_to_theta_msg_exact} reflects the uncertainty due to this unobservability through an infinitely broad, uninformative distribution for $\nt{\theta}$.

To avoid an improper pdf, we modify how this message is derived by regarding our assumed signal model, in which $s_n^{(t)} \in \{0, 1\}$, as a limiting case of the model with $s_n^{(t)} \in \{\eps, 1\}$ as $\eps \to 0$.  For any fixed positive $\eps$, the resulting message $\nu_{f_n^{(t)} \to \theta_n^{(t)}}(\cdot)$ is proper, given by
\ifthenelse{\boolean{ONE_COLUMN}}
{
\begin{equation}
	\msg{\nt{f}}{\nt{\theta}}^{\text{mod}}(\nt{\theta}) = (1 - \Omega(\nt{\bwd{\pi}})) \,\, \mathcal{CN}(\nt{\theta}; \tfrac{1}{\eps} \phi_{nt}^i, \tfrac{1}{\eps^2} c_t^i) + \Omega(\nt{\bwd{\pi}}) \,\, \mathcal{CN}(\nt{\theta}; \phi_{nt}^i, c_t^i),
	\label{eq:f_to_theta_msg_eps}
\end{equation}
}
{
\begin{eqnarray}
	\msg{\nt{f}}{\nt{\theta}}^{\text{mod}}(\nt{\theta}) &=& (1 - \Omega(\nt{\bwd{\pi}})) \,\, \mathcal{CN}(\nt{\theta}; \tfrac{1}{\eps} \phi_{nt}^i, \tfrac{1}{\eps^2} c_t^i) \nonumber \\
	&\quad& + \,\, \Omega(\nt{\bwd{\pi}}) \,\, \mathcal{CN}(\nt{\theta}; \phi_{nt}^i, c_t^i),
	\label{eq:f_to_theta_msg_eps}
\end{eqnarray}
}
where
\begin{equation}
	\Omega(\pi) \triangleq \frac{\eps^2 \pi}{(1 - \pi) + \eps^2 \pi}.
	\label{eq:omega_defn}
\end{equation}
The pdf in \eqref{f_to_theta_msg_eps} is that of a binary Gaussian mixture.  If we consider $\eps \ll 1$, the first mixture component is extremely broad, while the second is more ``informative,'' with mean $\phi_n^i$ and variance $c_n^i$.  The relative weight assigned to each component Gaussian is determined by the term $\Omega(\bwd{\pi}_n^{(t)})$.  Notice that the limit of this weighting term is the simple indicator function 
\begin{equation}
	\lim_{\eps \to 0} \Omega(\pi) = 
	\begin{cases}
		0 & \text{if } 0 \le \pi < 1, \\
		1 & \text{if } \pi = 1.
	\end{cases}
	\label{eq:omega_indicator}
\end{equation}

Since we cannot set $\eps = 0$, we instead fix a small positive value, e.g., $\eps = 10^{-7}$.  In this case, \eqref{f_to_theta_msg_eps} could then be used as the outgoing message.  However, this presents a further difficulty: propagating a binary Gaussian mixture forward in time would lead to an exponential growth in the number of mixture components at subsequent timesteps.  This difficulty is a familiar one in the context of switched linear dynamical systems based on conditional Gaussian models, since such models are not closed under marginalization \cite{ZH2005}.  To avoid the exponential growth in the number of mixture components, we collapse our binary Gaussian mixture to a single Gaussian component, an approach sometimes referred to as a Gaussian sum approximation \cite{AS1972,BC2010}.  This can be justified by the fact that, for $\eps \ll 1$, $\Omega(\cdot)$ behaves nearly like the indicator function in \eqref{omega_indicator}, in which case one of the two Gaussian components will typically have negligible mass.

To carry out the Gaussian sum approximation, we propose the following two schemes.  The first is to simply choose a threshold $\tau$ that is slightly smaller than $1$ and, using \eqref{omega_indicator} as a guide, threshold $\nt{\bwd{\pi}}$ to choose between the two Gaussian components of \eqref{f_to_theta_msg_eps}.  The resultant message is thus
\begin{equation}
	\nu_{f_n^{(t)} \to \theta_n^{(t)}}(\theta_n^{(t)}) = \mathcal{CN}(\theta_n^{(t)}; \fwd{\xi}_n^{(t)}, \fwd{\psi}_n^{(t)}),
	\label{eq:f_to_theta_msg}
\end{equation}
with $\fwd{\xi}_n^{(t)}$ and $\fwd{\psi}_n^{(t)}$ chosen according to
\begin{equation}
	\big(\fwd{\xi}_n^{(t)}, \fwd{\psi}_n^{(t)}\big) = \left\{
	\begin{array}{ll}
	\big(\tfrac{1}{\eps} \phi_n^i, \tfrac{1}{\eps^2} c_n^i\big), & \bwd{\pi}_n^{(t)} \le \tau \\
	\big(\phi_n^i, c_n^i\big), & \bwd{\pi}_n^{(t)} > \tau
	\end{array} \right. .
	\label{eq:amplitude_threshold}
\end{equation}
The second approach is to perform a second-order Taylor series approximation of $- \log \msg{\nt{f}}{\nt{\theta}}^{\text{mod}}(\nt{\theta})$ with respect to $\nt{\theta}$.  \textr{The resultant quadratic form in $\nt{\theta}$ can be viewed as the logarithm of a Gaussian kernel with a particular mean and variance, which can be used to parameterize} a single Gaussian message, as described in \cite{ZS2013}.  The latter approach has the advantage of being parameter-free.  Empirically, we find that this latter approach works well when changes in the support occur infrequently, e.g., $\pzo < 0.025$, while the former approach is better suited to more dynamic environments.

In \tabref{algorithm_equations} we provide a pseudo-code implementation of our proposed DCS-AMP algorithm that gives the explicit message update equations appropriate for performing a single forward pass.  \textr{The interested reader can find an expanded derivation of the messages in \cite{Z2013}.}  The primary computational burden of DCS-AMP is computing the messages passing between the $\{x_n^{(t)}\}$ and $\{g_m^{(t)}\}$ nodes, a task which can be performed efficiently using matrix-vector products involving $\vect{A}$ and $\vec{A}^{(t)^\textsf{H}}$.  The resulting overall complexity of DCS-AMP is therefore $\mathcal{O}(TMN)$ flops (flops-per-pass) when filtering (smoothing).\footnote{When they exist, fast implicit $\vect{A}$ operators can provide significant computational savings in high-dimensional problems.  Implementing a Fourier transform as a fast Fourier transform (FFT) subroutine, for example, would drop DCS-AMP's complexity from $\mathcal{O}(TMN)$ to $\mathcal{O}(TN \log_2 N)$.}  The storage requirements are $\mathcal{O}(N)$ and $\mathcal{O}(TN)$ complex numbers when filtering and smoothing, respectively.

\begin{table}[t]
\centering
\scriptsize
\setlength{\tabcolsep}{2pt}
\setlength{\belowcaptionskip}{3ex}
\begin{tabular}{|llrclr|}
	\hline
	\multicolumn{5}{|l}{$\textsf{\% Define soft-thresholding functions: }$}&\\
	& \multicolumn{4}{l}{$\textit{F}_{nt}(\phi; c) \triangleq (1 + \gamma_{nt}(\phi; c))^{-1}  
		\Big( \frac{\bwd{\psi}_n^{(t)} \phi + \bwd{\xi}_n^{(t)} c}{\bwd{\psi}_n^{(t)} + c} \Big)$}	& \threshcnt{eq:thresh_start}\\[2ex]
	& \multicolumn{4}{l}{$\textit{G}_{nt}(\phi; c) \triangleq (1 + \gamma_{nt}(\phi; c))^{-1} 
		\Big( \frac{\bwd{\psi}_n^{(t)} c}{\bwd{\psi}_n^{(t)} + c} \Big) + \gamma_{nt}(\phi; c) |\textit{F}_{nt}(\phi; c)|^2$}	& \threshcnt{}\\[1ex]
	& \multicolumn{4}{l}{$\textit{F}_{nt}'(\phi; c) \triangleq \tfrac{\partial}{\partial \phi} \textit{F}_{nt}(\phi,c) = \tfrac{1}{c} \textit{G}_{nt}(\phi; c)$}	& \threshcnt{}\\[1ex]
	& \multicolumn{4}{l}{$\gamma_{nt}(\phi; c) \triangleq \Big( \frac{1 - \bwd{\pi}_n^{(t)}}{\bwd{\pi}_n^{(t)}} \Big) \Big( \frac{\bwd{\psi}_n^{(t)} + c}{c} \Big) $} &\\
	&&& \multicolumn{2}{l}{$\quad \times \exp\Big( - \Big[ \frac{\bwd{\psi}_n^{(t)} |\phi|^2 + \bwd{\xi}_n^{(t)\,*} c \phi + 
		\bwd{\xi}_n^{(t)} c \phi^* - c |\bwd{\xi}_n^{(t)}|^2}{c(\bwd{\psi}_n^{(t)} + c)} \Big] \Big)$}	& \threshcnt{eq:thresh_end}\\[1ex]
	\hline
	\multicolumn{5}{|l}{$\textsf{\% Begin passing messages} \ldots$} &\\[-1ex]
  	\multicolumn{5}{|l}{$\textsf{for } t=1,\ldots,T:$}&\\[-1ex]
	&\multicolumn{4}{l}{$\quad \textsf{\% Execute the } \textbf{(into)} \textsf{ phase} \ldots$} &\\
	&\multicolumn{4}{l}{$\quad \nt{\bwd{\pi}} = \frac {\nt{\fwd{\lambda}} \cdot \nt{\bwd{\lambda}}} 
		{(1 - \nt{\fwd{\lambda}}) \cdot (1 - \nt{\bwd{\lambda}}) + \nt{\fwd{\lambda}} \cdot \nt{\bwd{\lambda}}} \quad \forall n$}	& \algcnt{} \\[2mm]
	&\multicolumn{4}{l}{$\quad \bwd{\psi}_{n}^{(t)} = \frac{\fwd{\kappa}_{n}^{(t)} \cdot \bwd{\kappa}_{n}^{(t)}}
		{\fwd{\kappa}_{n}^{(t)} + \bwd{\kappa}_{n}^{(t)}} \quad \forall n$}	& \algcnt{}\\[2mm]
	&\multicolumn{4}{l}{$\quad \bwd{\xi}_{n}^{(t)} = \bwd{\psi}_{n}^{(t)} \cdot \Big(\frac{\fwd{\eta}_{n}^{(t)}}
		{\fwd{\kappa}_{n}^{(t)}} + \frac{\bwd{\eta}_{n}^{(t)}}{\bwd{\kappa}_{n}^{(t)}}\Big) \quad \forall n$}	& \algcnt{}\\

	&\multicolumn{4}{l}{$\quad \textsf{\% Initialize AMP-related variables} \ldots$} &\\
	&\multicolumn{4}{l}{$\quad \forall m: z_{mt}^1 = y_{m}^{(t)}, \forall n: \mu_{nt}^1 = 0, \textsf{ and } c_t^1 = 100 \cdot \textstyle \sum_{n=1}^N \psi_n^{(t)}$}&\\

	&\multicolumn{5}{l|}{$\quad \textsf{\% Execute the } \textbf{(within)} \textsf{ phase using AMP} \ldots$} \\
  	&\multicolumn{4}{l}{$\quad \textsf{for $i=1,\ldots,I$, }:$}&\\
  	&&$\qquad \phi_{nt}^i$ &$=$& $\sum_{m=1}^M A_{mn}^{(t)\,*} z_{mt}^i + \mu_{nt}^i \quad \forall n$ & \algcnt{eq:amp_start}\\
	&&$\qquad \mu_{nt}^{i+1}$ &$=$& $\textit{F}_{nt}(\phi_{nt}^i; c_t^i) \quad \forall n$	&\algcnt{eq:amp_mu_defn}\\
	&&$\qquad v_{nt}^{i+1}$ &$=$& $\textit{G}_{nt}(\phi_{nt}^i; c_t^i) \quad \forall n$	& \algcnt{eq:amp_v_defn}\\
	&&$\qquad c_t^{i+1}$ &$=$& $\sigma_e^2 + \tfrac{1}{M} \sum_{n=1}^N v_{nt}^{i+1}$	& \algcnt{}\\
	&&$\qquad z_{mt}^{i+1}$ &$=$& $y_m^{(t)} - \vec{a}_{m}^{(t)\,\textsf{T}}\! \vec{\mu}_t^{i+1} + \tfrac{z_{mt}^i}{M} \sum_{n=1}^N \textit{F}_{nt}'(\phi_{nt}^i; c_t^i) \quad \forall m$	& \algcnt{eq:amp_end}\\
  	&\multicolumn{2}{l}{$\quad \textsf{end}$}&&&\\
	&\multicolumn{4}{l}{$\quad \hat{x}_n^{(t)} = \mu_{nt}^{I+1} \quad \forall n \qquad \textsf{\% Store current estimate of } x_n^{(t)}$}	& \algcnt{eq:amp_x_hat}\\

	&\multicolumn{5}{l|}{$\quad \textsf{\% Execute the } \textbf{(out)} \textsf{ phase} \ldots$} \\
	&\multicolumn{4}{l}{$\quad \fwd{\pi}_n^{(t)} = \Big(1 + \Big( \frac{\bwd{\pi}_n^{(t)}}{1 - \bwd{\pi}_n^{(t)}} \Big) \gamma_{nt}(\phi_{nt}^{I}; c_{t}^{I+1}) \Big)^{-1} \quad \forall n$}	& \algcnt{}\\[1.5ex]
	&\multicolumn{4}{l}{$\,\,\,\,(\fwd{\xi}_n^{(t)}, \fwd{\psi}_n^{(t)}) = \left\{
		\begin{array}{ll}
		(\phi_n^I/\eps, c_t^{I+1}/\eps^2), & \bwd{\pi}_n^{(t)} \le \tau \\
		(\phi_n^I, c_t^{I+1}), & \text{o.w.}
		\end{array} \right.  \quad \forall n \quad (\eps \ll 1)$}	& \algcnt{} \\[0ex]
	&\multicolumn{5}{l|}{$\quad \textsf{\% Execute the } \textbf{(across)} \textsf{ phase \textr{forward in time}} \ldots$} \\
	&\multicolumn{4}{l}{$\quad \fwd{\lambda}_n^{(t+1)} = \frac{\poz (1-\fwd{\lambda}_n^{(t)})(1 - \fwd{\pi}_n^{(t)}) + (1-\pzo) \fwd{\lambda}_n^{(t)} \fwd{\pi}_n^{(t)}} {(1-\fwd{\lambda}_n^{(t)})(1 - \fwd{\pi}_n^{(t)}) + \fwd{\lambda}_n^{(t)} \fwd{\pi}_n^{(t)}} \quad \forall n$}	& \algcnt{}\\[2ex]
	&\multicolumn{4}{l}{$\quad \fwd{\eta}_n^{(t+1)} = (1-\alpha) \Big(\frac{\fwd{\kappa}_{n}^{(t)} \fwd{\psi}_{n}^{(t)}}{\fwd{\kappa}_{n}^{(t)} + \fwd{\psi}_{n}^{(t)}}\Big) \Big(\frac{\fwd{\eta}_{n}^{(t)}}{\fwd{\kappa}_{n}^{(t)}} + \frac{\fwd{\xi}_{n}^{(t)}}{\fwd{\psi}_{n}^{(t)}}\Big) + \alpha \zeta \quad \forall n$}	& \algcnt{}\\[2ex]
	&\multicolumn{4}{l}{$\quad \fwd{\kappa}_{n}^{(t+1)} = (1-\alpha)^2 \Big(\frac{\fwd{\kappa}_{n}^{(t)} \fwd{\psi}_{n}^{(t)}}{\fwd{\kappa}_{n}^{(t)} + \fwd{\psi}_{n}^{(t)}}\Big) + \alpha^2 \rho \quad \forall n$}	& \algcnt{}\\
	\multicolumn{5}{|l}{$\textsf{end}$}&\\
	\hline

\end{tabular}

\caption{DCS-AMP steps for filtering mode, or the forward portion of a single forward/backward pass in smoothing mode.  \textr{See \figref{message_summary} to associate quantities with the messages traversing the factor graph.}}
\label{tab:algorithm_equations}
\end{table}

\section{Learning the signal model parameters}
\label{sec:em_updates}
The signal model of \secref{signal_model} is specified by the Markov chain parameters $\lambda$, $\pzo$, the Gauss-Markov parameters $\zeta$, $\alpha$, $\rho$, and the AWGN variance $\sigma_e^2$.  It is likely that some or all of these parameters will require tuning in order to best match the unknown signal.  To this end, we develop an expectation-maximization (EM) \cite{DLR1977} algorithm that works together with the message passing procedure described in \secref{algorithm:scheduling} to learn all of the model parameters in an iterative fashion from the data.

The EM algorithm is appealing for two principal reasons.  First, the EM algorithm is a well-studied and principled means of parameter estimation.  At every EM iteration, the likelihood is guaranteed to increase until convergence to a local maximum occurs \cite{M1996}.  For multimodal likelihoods, local maxima will, in general, not coincide with the global maximum, but a judicious initialization of parameters can help in ensuring the EM algorithm reaches the global maximum \cite{M1996}.  The second appealing feature of the EM algorithm lies in the fact that its expectation step leverages quantities that have already been computed in the process of executing DCS-AMP, making the EM procedure computationally efficient.

We let $\Gamma \triangleq \{\lambda, \pzo, \zeta, \alpha, \rho, \sigma_e^2\}$ denote the set of all model parameters, and let $\Gamma^k$ denote the set of parameter estimates at the $k^{th}$ EM iteration.  The objective of the EM procedure is to find parameter estimates that maximize the data likelihood $p(\ovec{y}|\Gamma)$.  Since it is often computationally intractable to perform this maximization, the EM algorithm incorporates additional ``hidden'' data and iterates between two steps: \textit{(i)} evaluating the conditional expectation of the log likelihood of the hidden data given the observed data, $\ovec{y}$, and the current estimates of the parameters, $\Gamma^k$, and \textit{(ii)} maximizing this expected log likelihood with respect to the model parameters.  For all parameters except the noise variance, $\sigma_e^2$, we use $\ovec{s}$ and $\ovec{\theta}$ as the hidden data, while for $\sigma_e^2$ we use $\ovec{x}$.

Before running DCS-AMP, the model parameters are initialized using any available prior knowledge.  If operating in smoothing mode, DCS-AMP performs an initial forward/backward pass, as described in \secref{algorithm:scheduling}.  Upon completing this first pass, estimates of the marginal posterior distributions are available for each of the underlying random variables.  Additionally, belief propagation can provide pairwise joint posterior distributions, e.g., $p\big(s_n^{(t)}, s_n^{(t-1)}|\ovec{y}\big)$, for any variable nodes connected by a common factor node \cite{B2006}.  With these marginal, and pairwise joint, posterior distributions, it is possible to produce closed-form solutions for performing steps  \textit{(i)} and  \textit{(ii)} above.  We adopt a Gauss-Seidel scheme, performing coordinate-wise maximization, e.g.,
\begin{equation}
	\lambda^{k+1} = \argmax_\lambda \text{E}_{\ovec{s},\ovec{\theta}|\ovec{y}}\left[\log p(\ovec{y}, \ovec{s}, \ovec{\theta}; \lambda, \Gamma^k \! \setminus \!\! \{\! \lambda^k \!\}) \big| \ovec{y}; \Gamma^k \right]. \nonumber
\end{equation}
The EM procedure is performed after each forward/backward pass, leading to a convergent sequence of parameter estimates.  If operating in filtering mode, the procedure is similar, however the EM procedure is run after each recovered timestep using only causally available posterior estimates.

In \tabref{em_updates}, we provide the EM update equations for each of the parameters of our signal model, assuming DCS-AMP is operating in smoothing mode.  \textr{A complete derivation of each update can be found in \cite{Z2013}.}

\begin{table}
\centering
\scriptsize
\setlength{\tabcolsep}{2pt}
\setlength{\belowcaptionskip}{2ex}
\begin{tabular}{|llr|}
	\hline
	\multicolumn{3}{| l |}{$\textsf{\% Define key quantities obtained from AMP-MMV at iteration } k \textsf{:}$}\\
	$\text{E}\big[s_n^{(t)} \big| \ovec{y}\big] = \frac{\big(\fwd{\lambda}_n^{(t)} \fwd{\pi}_n^{(t)} \bwd{\lambda}_n^{(t)}\big)}{\big(\fwd{\lambda}_n^{(t)} \fwd{\pi}_n^{(t)} \bwd{\lambda}_n^{(t)} + (1\!-\!\fwd{\lambda}_n^{(t)})(\!1-\!\fwd{\pi}_n^{(t)})(\!1-\!\bwd{\lambda}_n^{(t)})\big)}$ & & (Q1)\\[2ex]
	$\text{E}\big[s_n^{(t)} s_n^{(t-1)} \big| \ovec{y}\big] = p\big(s_n^{(t)}=1, s_n^{(t-1)}=1 \big| \ovec{y}\big)$ & & (Q2)\\[2ex]
	$\tilde{v}_n^{(t)} \triangleq \text{var}\{\theta_n^{(t)} | \ovec{y}\} = \left(\frac{1}{\fwd{\kappa}_n^{(t)}} + \frac{1}{\fwd{\psi}_n^{(t)}} + \frac{1}{\bwd{\kappa}_n^{(t)}} \right)^{-1}$ & & (Q3)\\[2ex]
	$\tilde{\mu}_n^{(t)} \triangleq \text{E}[\theta_n^{(t)} | \ovec{y}] = \tilde{v}_n^{(t)} \cdot  \left(\frac{\fwd{\eta}_n^{(t)}}{\fwd{\kappa}_n^{(t)}} + \frac{\fwd{\xi}_n^{(t)}}{\fwd{\psi}_n^{(t)}} + \frac{\bwd{\eta}_n^{(t)}}{\bwd{\kappa}_n^{(t)}} \right)$ & & (Q4)\\[2ex]
 	$v_n^{(t)} \triangleq \text{var}\big\{x_n^{(t)} \big| \ovec{y}\big\} \qquad \text{\% See \eqref{amp_v_defn} of \tabref{algorithm_equations}}$ & & \\
	$\mu_n^{(t)} \triangleq \text{E}\big[x_n^{(t)} \big| \ovec{y}\big] \qquad \quad \text{\% See \eqref{amp_mu_defn} of \tabref{algorithm_equations}}$ & & \\[1ex]
 	\hline
	\multicolumn{3}{| l |}{$\textsf{\% EM update equations:}$}\\
	$\lambda^{k+1} = \tfrac{1}{N} \sum_{n=1}^{N} \text{E}\big[s_n^{(1)} \big| \ovec{y}\big]$ & & \emcnt{} \\[1ex]

	$\pzo^{k+1} = \frac{\sum_{t=2}^{T} \sum_{n=1}^{N} \text{E}\big[s_n^{(t\!-\!1)} \big| \ovec{y}\big] - \text{E}\big[s_n^{(t)} s_n^{(t\!-\!1)} \big| \ovec{y}\big] } {\sum_{t=2}^{T} \sum_{n=1}^{N} \text{E}\big[s_n^{(t\!-\!1)} \big| \ovec{y}\big] }$ & & \emcnt{} \\[2ex]
	
	$\zeta^{k+1} = \left( \tfrac{N(T-1)}{\rho^k} + \tfrac{N}{(\sigma^2)^k} \right)^{-1} \Big( \tfrac{1}{(\sigma^2)^k} \sum_{n=1}^N \tilde{\mu}_n^{(1)}$ & & \\[1ex]
	$\qquad \quad + \sum_{t=2}^T \sum_{n=1}^N \tfrac{1}{\alpha^k \rho^k} \big(\nt{\tilde{\mu}} - (1 - \alpha^k) \tilde{\mu}_n^{(t-1)}\big) \Big)$ & & \emcnt{} \\[1ex]
	
	$\alpha^{k+1} = \tfrac{1}{4N(T-1)} \Big(\mf{b} - \sqrt{\mf{b}^2 + 8N(T-1) \mf{c}}\Big)$ & & \emcnt{} \\[1ex]
	\quad where: & & \\[0ex]
	$\quad \mf{b} \triangleq \tfrac{2}{\rho^k} \sum_{t=2}^T \sum_{n=1}^N \mf{Re}\big\{ E[{\nt{\theta}}^{*} \theta_n^{(t-1)} | \ovec{y}] \big\}$  & & \\[0ex]
	$\qquad \qquad - \mf{Re}\{(\nt{\tilde{\mu}} - \tilde{\mu}_n^{(t-1)})^{*} \zeta^k\} - \tilde{v}_n^{(t-1)} - |\tilde{\mu}_n^{(t-1)}|^2$ & & \\[0ex]
	$\quad \mf{c} \triangleq \tfrac{2}{\rho^k} \sum_{t=2}^T \sum_{n=1}^N \nt{\tilde{v}} + |\nt{\tilde{\mu}}|^2 + \tilde{v}_n^{(t-1)} + |\tilde{\mu}_n^{(t-1)}|^2$ & & \\[0ex]
	$\qquad \qquad - 2 \mf{Re}\big\{ E[{\nt{\theta}}^{*} \theta_n^{(t-1)} | \ovec{y}] \big\}$ & & \\[0ex]
	$\rho^{k+1} = \tfrac{1}{(\alpha^k)^2 N (T-1)} \sum_{t=2}^T \sum_{n=1}^N \nt{\tilde{v}} + |\nt{\tilde{\mu}}|^2$ & & \\[0ex]
	$\qquad \qquad  + (\alpha^k)^2 |\zeta^k|^2 - 2 (1 - \alpha^k) \mf{Re}\big\{ E[{\nt{\theta}}^{*} \theta_n^{(t-1)} | \ovec{y}] \big\}$ & & \\[0ex]
	$\qquad \qquad - 2 \alpha^k \mf{Re}\big\{ \tilde{\mu}_n^{(t)*} \zeta^k \big\} + 2 \alpha^k (1-\alpha^k) \mf{Re}\big\{ \tilde{\mu}_n^{(t-1)*} \zeta^k \big\}$ & & \\[0ex]
	$\qquad \qquad + (1 - \alpha^k) (\tilde{v}_n^{(t-1)} + |\tilde{\mu}_n^{(t-1)}|^2)$ & & \emcnt{} \\[0ex]
	$(\sigma_e^2)^{k+1} = \tfrac{1}{TM} \left( \sum_{t=1}^T \|\vect{y} - \vec{A} \vect{\mu}\|^2 + \vec{1}_N^T \vect{v} \right)$ & & \emcnt{} \\[0ex]
	\hline
\end{tabular}

\caption{EM update equations for the signal model parameters of \secref{signal_model}.}
\label{tab:em_updates}
\end{table}

\section{Incorporating Additional Structure}
\label{sec:structure}
In Sections \ref{sec:signal_model} - \ref{sec:em_updates} we described a signal model for the dynamic CS problem and summarized a message passing algorithm for making inferences under this model, while iteratively learning the model parameters via EM.  We also hinted that the model could be generalized to incorporate additional, or more complex, forms of structure.  In this section we will elaborate on this idea, and illustrate one such generalization.

Recall that, in \secref{signal_model}, we introduced hidden variables $\ovec{s}$ and $\ovec{\theta}$ in order to characterize the structure in the signal coefficients.  An important consequence of introducing these hidden variables was that they made each signal coefficient $\nt{x}$ conditionally independent of the remaining coefficients in $\ovec{x}$, given $\nt{s}$ and $\nt{\theta}$.  This conditional independence served an important algorithmic purpose since it allowed us to apply the AMP algorithm, which requires independent local priors, within our larger inference procedure.

One way to incorporate additional structure into the signal model of \secref{signal_model} is to generalize our choices of $p(\ovec{s})$ and $p(\ovec{\theta})$.  \textr{As a concrete example, pairing the temporal support model proposed in this work with the Markovian  model of wavelet tree inter-scale correlations described in \cite{SS2012} through a more complex support prior, $p(\ovec{s})$, could enable even greater undersampling in a dynamic MRI setting.  Performing inference on such models could be accomplished through the general algorithmic framework proposed in \cite{ZRS2012}.}  As another example, suppose that we wish to expand our Bernoulli-Gaussian signal model to one in which signal coefficients are marginally distributed according to a Bernoulli-Gaussian-mixture, i.e.,
\begin{equation}
	p(\nt{x}) = \lambda_{0}^{(t)} \delta(\nt{x}) + \sum_{d=1}^D \lambda_{d}^{(t)} \mathcal{CN}(\nt{x}; \zeta_d, \sigma_d^2), \nonumber
	\label{eq:bernoulli_gauss_mixture}
\end{equation}
where $\sum_{d=0}^D \lambda_{d}^{(t)} = 1$.  Since we still wish to preserve the slow time-variations in the support and smooth evolution of non-zero amplitudes, a natural choice of hidden variables is $\{\ovec{s}, \ovec{\theta}_1, \ldots, \ovec{\theta}_D\}$, where $\nt{s} \in \{0,1,\ldots,D\}$, and $\theta_{d,n}^{(t)} \in \mathbb{C}, \, d = 1, \ldots, D$.  The relationship between $\nt{x}$ and the hidden variables then generalizes to:
\begin{equation}
	p(\nt{x} | \nt{s}, \theta_{1,n}^{(t)}, \ldots, \theta_{D,n}^{(t)}) = \left\{
	\begin{array}{cc}
		\delta(\nt{x}), & \nt{s} = 0, \\
		\delta(\nt{x} - \theta_{d,n}^{(t)}), & \nt{s} = d \neq 0. \\
	\end{array}
	\right.  \nonumber
	\label{eq:gauss_mix_cond_prior}
\end{equation}

To model the slowly changing support, we specify $p(\ovec{s})$ using a $(D+1)$-state Markov chain defined by the transition probabilities $p_{0d} \triangleq \text{Pr}\{\nt{s} = 0 | s_n^{(t-1)} = d\}$ and $p_{d0} \triangleq \text{Pr}\{\nt{s} = d | s_n^{(t-1)} = 0\}$, $d = 1, \ldots, D$.  In this work, we assume that state transitions cannot occur between active mixture components, i.e., $\text{Pr}(\nt{s} = d | s_n^{(t-1)} = e) = 0$ when $d \neq e \neq 0$.\footnote{\textr{By relaxing this restriction on active-to-active state transitions, we can model signals whose coefficients tend to enter the support set at small amplitudes that grow larger over time through the use of a Gaussian mixture component with a small variance that has a high probability of transitioning to a higher variance mixture component. \label{gm_foot}}}  For each amplitude time-series we again use independent Gauss-Markov processes to model smooth evolutions in the amplitudes of active signal coefficients, i.e.,
\begin{equation}
	\theta_{d,n}^{(t)} = (1-\alpha_d) \big(\theta_{d,n}^{(t-1)} - \zeta_d \big) + \alpha_d w_{d,n}^{(t)} + \zeta_d, \nonumber
	\label{eq:gm_theta_evolve}
\end{equation}
where $w_{d,n}^{(t)} \sim \mathcal{CN}(0, \rho_d)$.

As a consequence of this generalized signal model, a number of the message computations of \secref{algorithm:implementation} must be modified.  For steps $\textbf{(into)}$ and $\textbf{(across)}$, it is largely straightforward to extend the computations to account for the additional hidden variables.  For step $\textbf{(within)}$, the modifications will affect the AMP thresholding equations defined in \eqref{thresh_start} - \eqref{thresh_end} of \tabref{algorithm_equations}.  Details on a Bernoulli-Gaussian-mixture AMP algorithm can be found in \cite{VS2012}.  For the \textbf{(out)} step, we will encounter difficulties applying standard sum-product update rules to compute the messages $\{\msg{\nt{f}}{\theta_{d,n}^{(t)}}(\cdot)\}_{d=1}^D$.  As in the Bernoulli-Gaussian case, we consider a modification of our assumed signal model that incorporates an $\eps \ll 1$ term, and use Taylor series approximations of the resultant messages to collapse a $(D+1)$-ary Gaussian mixture to a single Gaussian.  More information on this procedure can be found in \cite{Z2013}.

\section{Empirical Study}
\label{sec:study}
We now describe the results of an empirical study of \textr{DCS-AMP}.\footnote{Code for reproducing our results is available at \url{http://www.ece.osu.edu/~schniter/turboAMPdcs}.}  The primary performance metric that we used in all of our experiments, which we refer to as the time-averaged normalized MSE (TNMSE), is defined as
\begin{equation}
	\text{TNMSE}(\ovec{x}, \hat{\ovec{x}}) \triangleq \frac{1}{T} \sum_{t=1}^T \frac{\|\vect{x} - \hvec{x}^{(t)}\|_2^2}{\|\vect{x}\|_2^2}, \nonumber
\end{equation}
where $\hvec{x}^{(t)}$ is an estimate of $\vect{x}$.

Unless otherwise noted, the following settings were used for DCS-AMP in our experiments.  First, DCS-AMP was run as a smoother, with a total of $5$ forward/backward passes.  The number of inner AMP iterations $I$ for each \textbf{(within)} step was $I = 25$, with a possibility for early termination if the change in the estimated signal, $\vec{\mu}_t^i$, fell below a predefined threshold from one iteration to the next, i.e., $\tfrac{1}{N} \|\vec{\mu}_t^i - \vec{\mu}_t^{i-1}\|_2 < 10^{-5}$.  \Eqref{amp_x_hat} of \tabref{algorithm_equations} was used to produce $\hvec{x}^{(t)}$, which corresponds to an MMSE estimate of $\vect{x}$ under DCS-AMP's estimated posteriors $p(\nt{x}|\ovec{y})$.  The amplitude approximation parameter $\eps$ from \eqref{f_to_theta_msg_eps} was set to $\eps = 10^{-7}$, while the threshold $\tau$ from \eqref{amplitude_threshold} was set to $\tau = 0.99$.  In our experiments, we found DCS-AMP's performance to be relatively insensitive to the value of $\eps$ provided that $\eps \ll 1$.  The choice of $\tau$ should be selected to provide a balance between allowing DCS-AMP to track amplitude evolutions on signals with rapidly varying supports and preventing DCS-AMP from prematurely gaining too much confidence in its estimate of the support.  We found that the choice $\tau = 0.99$ works well over a broad range of problems.  When the estimated transition probability $\pzo < 0.025$, DCS-AMP automatically switched from the threshold method to the Taylor series method of computing \eqref{f_to_theta_msg}, which is advantageous because it is parameter-free.

\textr{When learning model parameters adaptively from the data using the EM updates of \tabref{em_updates}, it is necessary to first initialize the parameters at reasonable values.  Unless domain-specific knowledge suggests a particular initialization strategy, we advocate using the following simple heuristics: The initial sparsity rate, $\lambda^1$, active mean, $\zeta^1$, active variance, $(\sigma^2)^1$, and noise variance, $(\sigma_e^2)^1$, can be initialized according to the procedure described in \cite[\S V]{VS2012}.\footnote{\textr{For problems with a high degree of undersampling and relatively non-sparse signals, it may be necessary to threshold the value for $\lambda^1$ suggested in \cite{VS2012} so that it does not fall below, e.g., $0.10$.}}  The Gauss-Markov correlation parameter, $\alpha$, can be initialized as
\begin{equation}
	\alpha^1 = 1 - \frac{1}{T-1} \sum_{t=1}^{T-1} \frac{|\vec{y}^{(t)\,\textsf{H}} \vec{y}^{(t+1)}|}{\lambda^1 (\sigma^2)^1 |\tr\{\vect{A} \vec{A}^{(t+1)\,\textsf{H}}\}|}.
\end{equation}
The active-to-inactive transition probability, $\pzo$, is difficult to gauge solely from sample statistics involving the available measurements, $\ovec{y}$.  We used $\pzo^1 = 0.10$ as a generic default choice, based on the premise that it is easier for DCS-AMP to adjust to more dynamic signals once it has a decent ``lock'' on the static elements of the support, than it is for it to estimate relatively static signals under an assumption of high dynamicity.}

\subsection{Performance across the sparsity-undersampling plane}
Two factors that have a significant effect on the performance of any CS algorithm are the sparsity $|\mathcal{S}^{(t)}|$ of the underlying signal, and the number of measurements $M$.  Consequently, much can be learned about an algorithm by manipulating these factors and observing the resulting change in performance.  To this end, we studied DCS-AMP's performance across the sparsity-undersampling plane \cite{DT2009}, which is parameterized by two quantities, the \emph{normalized sparsity ratio}, $\beta \triangleq \text{E}[|\mathcal{S}^{(t)}|] / M$, and the \emph{undersampling ratio}, $\delta \triangleq M / N$.  For a given $(\delta, \beta)$ pair (with $N$ fixed at $1500$), sample realizations of $\ovec{s}$, $\ovec{\theta}$, and $\ovec{e}$ were drawn from their respective priors, and elements of a time-varying $\vec{A}^{(t)}$ were drawn from i.i.d. zero-mean complex circular Gaussians, with all columns subsequently scaled to have unit $\ell_2$-norm, thus generating $\ovec{x}$ and $\ovec{y}$.

As a performance benchmark, we used the support-aware Kalman smoother.  In the case of linear dynamical systems with jointly Gaussian signal and observations, the Kalman filter (smoother) is known to provide MSE-optimal causal (non-causal) signal estimates \cite{E2006}.  When the signal is Bernoulli-Gaussian, the Kalman filter/smoother is no longer optimal.  However, a lower bound on the achievable MSE can be obtained using the support-aware Kalman filter (SKF) or smoother (SKS).  Since the classical state-space formulation of the Kalman filter does not easily yield the support-aware bound, we turn to an alternative view of Kalman filtering as an instance of message passing on an appropriate factor graph \cite{LDHKPK2007}.  For this, it suffices to use the factor graph of \figref{total_factor_graph} with $\{s_n^{(t)}\}$ treated as fixed, known quantities.  Following the standard sum-product algorithm rules results in a message passing algorithm in which all messages are Gaussian, and no message approximations are required.  Then, by running loopy Gaussian belief propagation until convergence, we are guaranteed that the resultant posterior means constitute the MMSE estimate of $\ovec{x}$ \cite[Claim 5]{WF2001}.

To quantify the improvement obtained by exploiting temporal correlation, signal recovery was also explored using the Bernoulli-Gaussian AMP algorithm (BG-AMP) independently at each timestep (i.e., ignoring temporal structure in the support and amplitudes), accomplished by passing messages only within the dashed boxes of \figref{total_factor_graph} using $p\big(x_n^{(t)}\big)$ from \eqref{x_prior} as AMP's prior.\footnote{Experiments were also run that compared performance against Basis Pursuit Denoising (BPDN) \cite{CDS1998} with genie-aided parameter tuning (solved using the SPGL1 solver \cite{BF2008}).  However, this was found to yield higher TNMSE than BG-AMP, and at higher computational cost.}

\begin{figure*}
	\begin{center}
		\ifthenelse{\boolean{ONE_COLUMN}}
		{\scalebox{0.37}{\includegraphics*[-4.5in,3.15in][12.95in,7.95in]{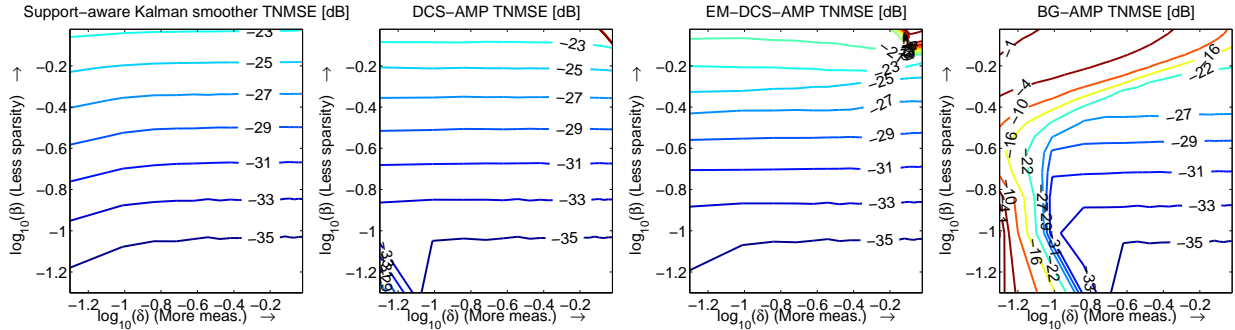}}}
		{\scalebox{0.39}{\includegraphics*[-4.5in,3.15in][12.95in,7.95in]{07Apr2013_182813165_phase_plane}}}
		\caption{A plot of the TNMSE (in dB) of (from left) the SKS, DCS-AMP, EM-DCS-AMP, and BG-AMP across the sparsity-undersampling plane, for temporal correlation parameters $\pzo = 0.05$ and $\alpha = 0.01$.}
		\label{fig:phase_plane_TNMSE}
	\end{center}
\end{figure*}

In \figref{phase_plane_TNMSE}, we present four plots from a representative experiment.  \textr{The TNMSE across the (logarithmically scaled) sparsity-undersampling plane is shown for (working from left to right) the SKS, DCS-AMP, EM-DCS-AMP (DCS-AMP with EM parameter tuning), and BG-AMP.  In order to allow EM-DCS-AMP ample opportunity to converge to the correct parameter values, it was allowed up to $300$ EM iterations/smoothing passes, although it would quite often terminate much sooner if the parameter initializations were reasonably close.}  The results shown were averaged over more than \textr{$300$} independent trials at each $(\delta,\beta)$ pair.   For this experiment, signal model parameters were set at \textr{$N = 1500$}, $T = 25$, $\pzo = 0.05$, $\zeta = 0$, $\alpha = 0.01$, $\sigma^2 = 1$, and a noise variance, $\sigma_e^2$, chosen to yield a signal-to-noise ratio (SNR) of \textr{$25$} dB.  $(M, \lambda)$ were set based on specific $(\delta, \beta)$ pairs, and $\poz$ was set so as to keep the expected number of active coefficients constant across time.  It is interesting to observe that the performance of the SKS and (EM-)DCS-AMP are only weakly dependent on the undersampling ratio $\delta$.  In contrast, the structure-agnostic BG-AMP algorithm is strongly affected.  This is one of the principal benefits of incorporating temporal structure; it makes it possible to tolerate more substantial amounts of undersampling, particularly when the underlying signal is less sparse.

\subsection{Performance vs $\pzo$ and $\alpha$}
The temporal correlation of our time-varying sparse signal model is largely dictated by two parameters, the support transition probability $\pzo$ and the amplitude forgetting factor $\alpha$.  Therefore, it is worth investigating how the performance of (EM-)DCS-AMP is affected by these two parameters.  In an experiment similar to that of \figref{phase_plane_TNMSE}, we tracked the performance of (EM-)DCS-AMP, the SKS, and BG-AMP across a plane of $(\pzo,\alpha)$ pairs.  The active-to-inactive transition probability $\pzo$ was swept linearly over the range $[0, 0.15]$, while the Gauss-Markov amplitude forgetting factor $\alpha$ was swept logarithmically over the range $[0.001, 0.95]$.  To help interpret the meaning of these parameters, we note that the fraction of the support that is expected to change from one timestep to the next is given by $2 \,\pzo$, and that the Pearson correlation coefficient between temporally adjacent amplitude variables is $1 - \alpha$.

\begin{figure*}
	\begin{center}
		\ifthenelse{\boolean{ONE_COLUMN}}
		{\scalebox{0.41}{\includegraphics*[-3in,3.00in][12in,8.0in]{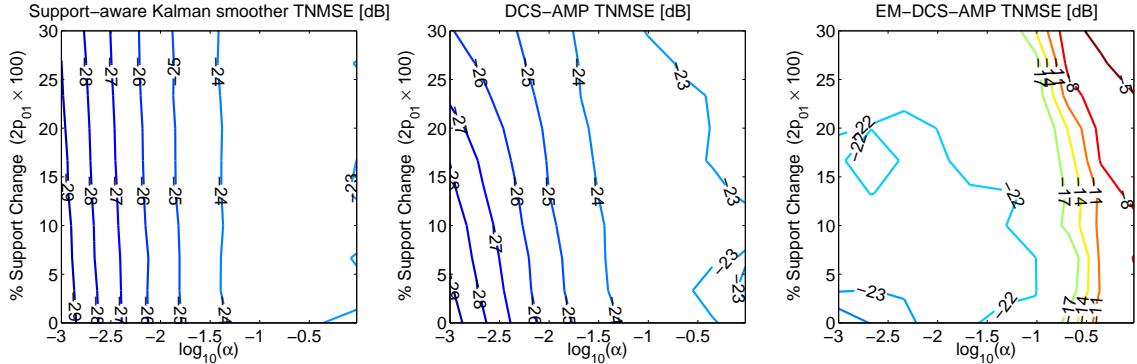}}}
		{\scalebox{0.38}{\includegraphics*[-3in,3.00in][12in,8.0in]{12Apr2013_193857522_dynamics}}}
		\caption{TNMSE (in dB) of (from left) the SKS, DCS-AMP, and EM-DCS-AMP as a function of the model parameters $\pzo$ and $\alpha$, for undersampling ratio $\delta = 1/3$ and sparsity ratio $\beta = 0.45$.  BG-AMP achieved a TNMSE of \textr{$-5.9$} dB across the plane.}
		\label{fig:dynamics_TNMSE_diff}
	\end{center}
\end{figure*}

In \figref{dynamics_TNMSE_diff} we plot the TNMSE (in dB) of the SKS and \mbox{(EM-)DCS-AMP} as a function of the percentage of the support that changes from one timestep to the next (i.e., $2 \pzo \times 100$) and the logarithmic value of $\alpha$ for a signal model in which $\delta = 1/5$ and $\beta = 0.60$, with remaining parameters set as before.  Since BG-AMP is agnostic to temporal correlation, its performance is insensitive to the values of $\pzo$ and $\alpha$.  Therefore, we do not include a plot of the performance of BG-AMP, but note that it achieved a TNMSE of \textr{$-5.9$} dB across the plane.  For the SKS and (EM-)DCS-AMP, we see that performance improves with increasing amplitude correlation (moving leftward).  This behavior, although intuitive, is in contrast to the relationship between performance and correlation found in the MMV problem \cite{ZR2011,ZS2013}, primarily due to the fact that the measurement matrix is static for all timesteps in the classical MMV problem, whereas here it varies with time.  Since the SKS has perfect knowledge of the support, its performance is only weakly dependent on the rate of support change.  DCS-AMP performance shows a modest dependence on the rate of support change, but nevertheless is capable of managing rapid temporal changes in support, while EM-DCS-AMP performs very near the level of the noise when $\alpha < 0.10$.

\subsection{Recovery of an MRI image sequence}
\label{sec:study:mri}
While the above simulations demonstrate the effectiveness of DCS-AMP in recovering signals generated according to our signal model, it remains to be seen whether the signal model itself is suitable for describing practical dynamic CS signals.  To address this question, we tested the performance of DCS-AMP on a dynamic MRI experiment first performed in \cite{LV2009}.  The experiment consists of recovering a sequence of $10$ MRI images of the larynx, each $256 \times 256$ pixels in dimension.  (See \figref{larynx_recoveries}.)  The measurement matrices were never stored explicitly due to the prohibitive sizes involves, but were instead treated as the composition of three linear operations, $\vec{A} = \vec{MFW}^{\textsf{T}}$.  The first operation, $\vec{W^{\textsf{T}}}$, was the synthesis of the underlying image from a sparsifying 2-D, 2-level Daubechies-4 wavelet transform representation.  The second operation, $\vec{F}$, was a 2-D Fourier transform that yielded the k-space coefficients of the image.  The third operation, $\vec{M}$, was a sub-sampling mask that kept only a fraction of the available k-space data.

Since the image transform coefficients are compressible rather than sparse, the SKF/SKS no longer serves as an appropriate algorithmic benchmark.  Instead, we compare performance against Modified-CS \cite{VL2010}, as well as timestep-independent Basis Pursuit.\footnote{Modified-CS is available at \url{http://home.engineering.iastate.edu/~luwei/modcs/index.html}.  Basis Pursuit was solved using the $\ell_1$-MAGIC equality-constrained primal-dual solver (chosen since it is used as a subroutine within Modified-CS), available at \url{http://users.ece.gatech.edu/~justin/l1magic/}.}  As reported in \cite{VL2010}, Modified-CS demonstrates that substantial improvements can be obtained over temporally agnostic methods.

Since the statistics of wavelet coefficients at different scales are often highly dissimilar (e.g., the coarsest-scale approximation coefficients are usually much less sparse than those at finer scales, and are also substantially larger in magnitude), we allowed our EM procedure to learn different parameters for different wavelet scales.  Using $\mathcal{G}_1$ to denote the indices of the coarsest-scale ``approximation'' coefficients, and $\mathcal{G}_2$ to denote the finer-scale ``wavelet'' coefficients, DCS-AMP was initialized with the following parameter choices: $\lambda_{\mathcal{G}_1} = 0.99$, $\lambda_{\mathcal{G}_2} = 0.01$, $\pzo = 0.01$, $\zeta_{\mathcal{G}_1} = \zeta_{\mathcal{G}_2} = 0$, $\alpha_{\mathcal{G}_1} = \alpha_{\mathcal{G}_2} = 0.05$, $\rho_{\mathcal{G}_1} = 10^5$, $\rho_{\mathcal{G}_2} = 10^4$, and $\sigma_e^2 = 0.01$, and run in filtering mode with $I = 10$ inner AMP iterations.

\textr{We note that our initializations were deliberately chosen to be agnostic, but reasonable, values.  In particular, observing that the coarsest-scale approximation coefficients of a wavelet decomposition are almost surely non-zero, we initialized the associated group's sparsity rate at $\lambda_{\mathcal{G}_1} = 0.99$, while the finer scale detail coefficients were given an arbitrary sparsity-promoting rate of $\lambda_{\mathcal{G}_2} = 0.01$.  The choices of $\alpha$ and $\rho$ were driven by an observation that the variance of coefficients across wavelet scales often differs by an order-of-magnitude.  The noise variance is arguably the most important parameter to initialize properly, since it balances the conflicting objectives of fitting the data and adhering to the assumed signal model.  Our rule-of-thumb for initializing this parameter was that it is best to err on the side of fitting the data (since the SNR in this MRI data collection was high), and thus we initialized the noise variance with a small value.}

In \tabref{MRI_table1} we summarize the performance of three different estimators: timestep-independent Basis Pursuit, which performs independent $\ell_1$ minimizations at each timestep, Modified-CS, and DCS-AMP (operating in filtering mode).  In this experiment, per the setup described in \cite{LV2009}, the initial timestep was sampled at $50\%$ of the Nyquist rate, i.e., $M = N/2$, while subsequent timesteps were sampled at $16\%$ of the Nyquist rate.  Both Modified-CS and DCS-AMP substantially outperform Basis Pursuit with respect to TNMSE, with DCS-AMP showing a slight advantage over Modified-CS.  Despite the similar TNMSE performance, note that DCS-AMP runs in seconds, whereas Modified-CS takes multiple hours.  In \figref{larynx_recoveries}, we plot the true images along with the recoveries for this experiment, which show severe degradation for Basis Pursuit on all but the initial timestep.
\begin{table}
	\centering
	\begin{tabular}{| c | c | c |}
		\hline
		\textbf{Algorithm}	&	\textbf{TNMSE (dB)}	&	\textbf{Runtime} \\
		\hline
		Basis Pursuit	&	-17.22	&	47 min	\\ \hline
		Modified-CS	&	-34.30	&	7.39 hrs	\\ \hline
		DCS-AMP (Filter)	&	\textbf{-34.62}	&	\textbf{8.08 sec}	\\ \hline
	\end{tabular}
	\caption{Performance on dynamic MRI dataset from \cite{LV2009} with increased sampling rate at initial timestep.}
	\label{tab:MRI_table1}
\end{table}
\ifthenelse{\boolean{ONE_COLUMN}}
{
\begin{figure*}
	\begin{center}
		\scalebox{0.35}{\includegraphics*[-1.3in,0.90in][9.95in,9.90in]{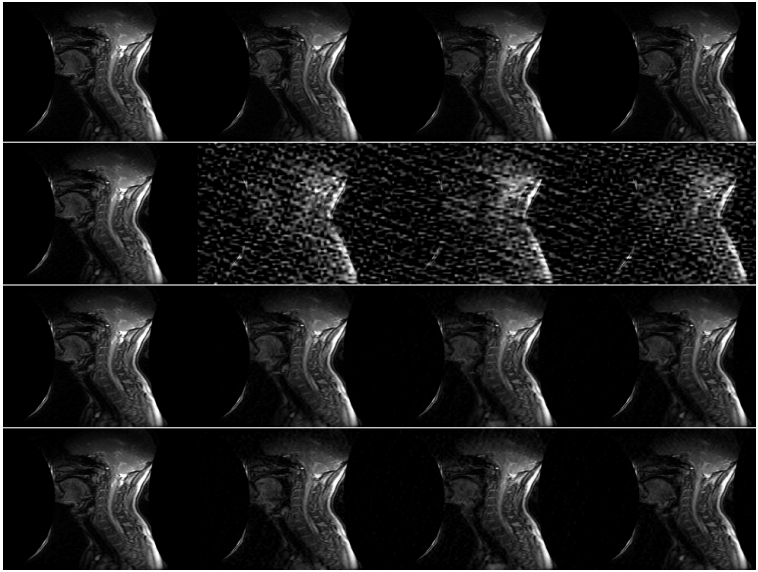}}
		\caption{Frames $1$, $2$, $5$, and $10$ of the dynamic MRI image sequence of (from top to bottom): the fully sampled dataset, Basis Pursuit, Modified-CS, and DCS-AMP, with increased sampling rate at initial timestep.}
		\label{fig:larynx_recoveries}
	\end{center}
\end{figure*}
}
{
\begin{figure}
	\begin{center}
		\scalebox{0.25}{\includegraphics*[-1.3in,0.90in][9.95in,9.90in]{larynx_recoveries_seq}}
		\caption{Frames $1$, $2$, $5$, and $10$ of the dynamic MRI image sequence of (from top to bottom): the fully sampled dataset, Basis Pursuit, Modified-CS, and DCS-AMP, with increased sampling rate at initial timestep.}
		\label{fig:larynx_recoveries}
	\end{center}
\end{figure}
}

In practice, it may not be possible to acquire an increased number of samples at the initial timestep.  We therefore repeated the experiment while sampling at $16\%$ of the Nyquist rate at every timestep.  The results, shown in \tabref{MRI_table2}, show that DCS-AMP's performance degrades by about $5$ dB, while Modified-CS suffers a $14$ dB reduction, illustrating that, when the estimate of the initial support is poor, Modified-CS struggles to outperform Basis Pursuit.
\begin{table}
	\centering
	\begin{tabular}{| c | c | c |}
		\hline                                                                                                                                                                                                                       
		\textbf{Algorithm}	&	\textbf{TNMSE (dB)}	&	\textbf{Runtime} \\
		\hline
		Basis Pursuit	&	-16.83	&	47.61 min	\\ \hline
		Modified-CS	&	-17.18	&	7.78 hrs	\\ \hline
		DCS-AMP (Filter)	&	\textbf{-29.51}	&	\textbf{7.27 sec}	\\ \hline
	\end{tabular}
	\caption{Performance on dynamic MRI dataset from \cite{LV2009} with identical sampling rate at every timestep.}
	\label{tab:MRI_table2}
\end{table}

\subsection{Recovery of a CS audio sequence}
In another experiment using real-world data, we used DCS-AMP to recover an audio signal from sub-Nyquist samples.  In this case, we employ the Bernoulli-Gaussian-mixture signal model proposed for DCS-AMP in \secref{structure}.  The audio clip is a $7$ second recording of a trumpet solo, and contains a succession of rapid changes in the trumpet's pitch.  Such a recording presents a challenge for CS methods, since the signal will be only compressible, and not sparse.  The clip, sampled at a rate of $11$ kHz, was divided into $T = 54$ non-overlapping segments of length $N = 1500$.  Using the discrete cosine transform (DCT) as a sparsifying basis, linear measurements were obtained using a time-invariant i.i.d. Gaussian sensing matrix.
 
\begin{figure}
	\begin{center}
		\scalebox{0.45}{\includegraphics*[0.35in,2.70in][7.75in,8.25in]{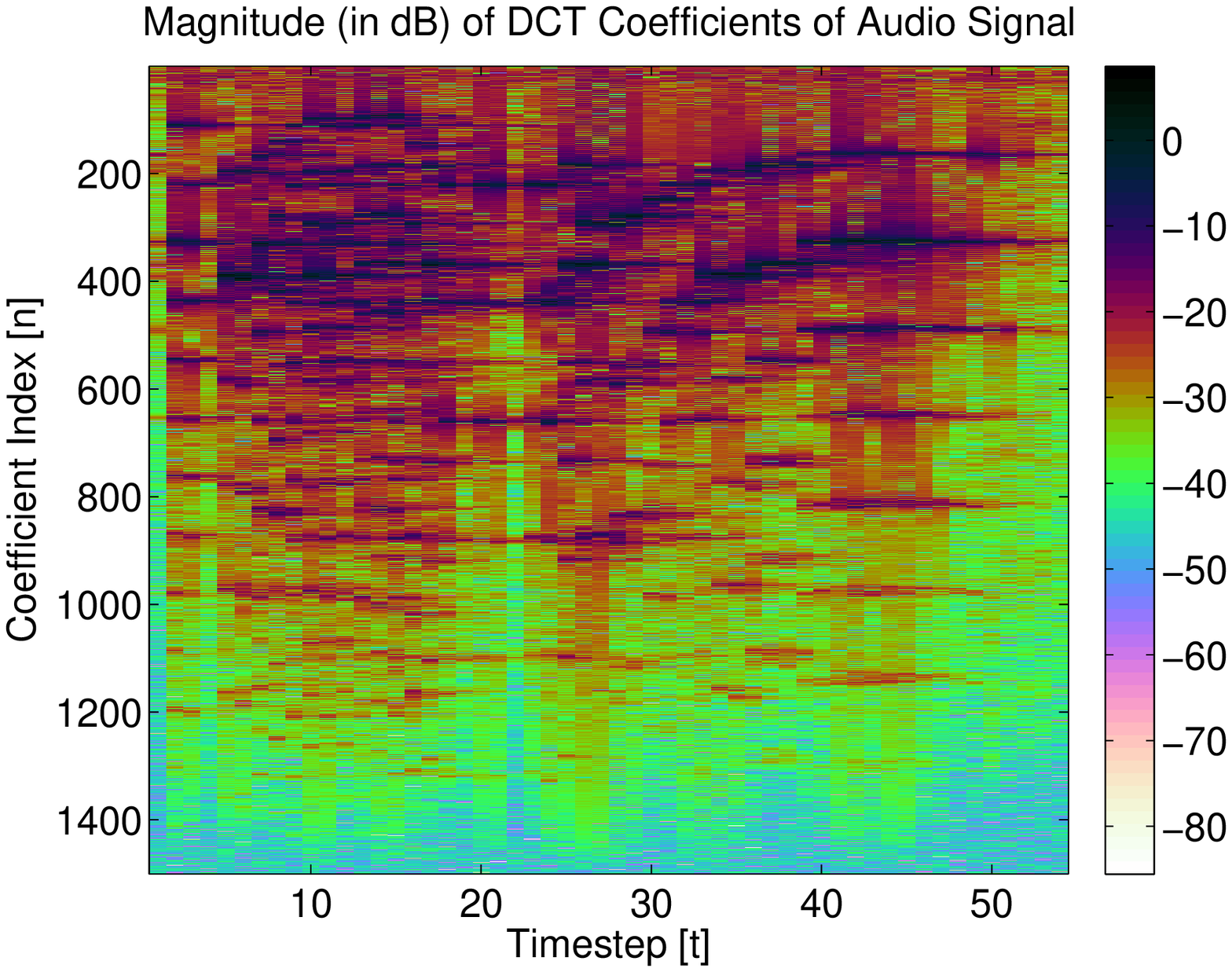}}
		\caption{DCT coefficient magnitudes (in dB) of an audio signal.}
		\label{fig:audio_dct}
	\end{center}
\end{figure}

 In \figref{audio_dct} we plot the magnitude of the DCT coefficients of the audio signal on a dB scale.  Beyond the temporal correlation evident in the plot, it is also interesting to observe that there is a non-trivial amount of frequency correlation (correlation across the index $[n]$), as well as a large dynamic range.  We performed recoveries using four techniques: BG-AMP, GM-AMP (a temporally agnostic Bernoulli-Gaussian-mixture AMP algorithm with $D=4$ Gaussian mixture components), DCS-(BG)-AMP, and DCS-GM-AMP (the Bernoulli-Gaussian-mixture dynamic CS model described in \secref{structure}, with $D = 4$).  For each algorithm, EM learning of the model parameters was performed using straightforward variations of the procedure described in \secref{em_updates}, with model parameters initialized automatically using simple heuristics described in \cite{VS2012}.  Moreover, unique model parameters were learned at each timestep (with the exception of support transition probabilities).  Furthermore, since our model of hidden amplitude evolutions was poorly matched to this audio signal, we fixed $\alpha = 1$.

In \tabref{audio_cs} we present the results of applying each algorithm to the audio dataset for three different undersampling rates, $\delta$.  For each algorithm, both the TNMSE in dB and the runtime in seconds are provided.  Overall, we see that performance improves at each undersampling rate as the signal model becomes more expressive.  While GM-AMP outperforms BG-AMP at all undersampling rates, it is surpassed by DCS-BG-AMP and DCS-GM-AMP, with DCS-GM-AMP offering the best TNMSE performance.  Indeed, we observe that one can obtain comparable, or even better, performance with an undersampling rate $\delta = \tfrac{1}{5}$ using DCS-BG-AMP or DCS-GM-AMP, with that obtained using BG-AMP with an undersampling rate $\delta = \tfrac{1}{3}$.
\begin{table*}
	\centering
	\ifthenelse{\boolean{ONE_COLUMN}}
	{
	\begin{tabular}{| c | c | c | c | c |}
		\hline
		\multicolumn{2}{|c|}{} & \multicolumn{3}{c|}{\textbf{Undersampling Rate}} \\ \cline{3-5}
		\multicolumn{2}{|c|}{} & \textbf{$\delta = \tfrac{1}{2}$}	&	\textbf{$\delta = \tfrac{1}{3}$}	&	\textbf{$\delta = \tfrac{1}{5}$} \\
		\hline
		\multirow{4}{*}{\rotatebox{90}{\textbf{Algorithm\,\,\,\,\,\,}}} & BG-AMP	&	-16.88 (dB) $\mid$ \textbf{09.11 (s)}&	-11.67 (dB) $\mid$ \textbf{08.27 (s)}	& -08.56 (dB) $\mid$ \textbf{06.63 (s)} \\ \cline{2-5}
		& GM-AMP ($D = 4$)	&	-17.49 (dB) $\mid$ 19.36 (s)	&	-13.74 (dB) $\mid$ 17.48 (s)	& -10.23 (dB) $\mid$ 15.98 (s) \\ \cline{2-5}
		& DCS-BG-AMP	&	-19.84 (dB) $\mid$ 10.20 (s)	&	-14.33 (dB) $\mid$ 08.39 (s)	& -11.40 (dB) $\mid$ 6.71 (s) \\ \cline{2-5}
		& DCS-GM-AMP ($D = 4$)	&	\textbf{-21.33 (dB)} $\mid$ 20.34 (s)	&	\textbf{-16.78 (dB)} $\mid$ 18.63 (s)	& \textbf{-12.49 (dB)} $\mid$ 10.13 (s) \\ \cline{1-5}
	\end{tabular}
	}
	{
	\begin{tabular}{| >{\centering\arraybackslash}m{0.25cm} | >{\centering\arraybackslash}m{3cm} | >{\centering\arraybackslash}m{3cm} | >{\centering\arraybackslash}m{3cm} | >{\centering\arraybackslash}m{3cm} |}
		\hline
		\multicolumn{2}{|c|}{} & \multicolumn{3}{c|}{\textbf{Undersampling Rate}} \\ \cline{3-5}
		\multicolumn{2}{|c|}{} & \textbf{$\delta = \tfrac{1}{2}$}	&	\textbf{$\delta = \tfrac{1}{3}$}	&	\textbf{$\delta = \tfrac{1}{5}$} \\[3 pt]
		\hline
		\multirow{4}{*}{\rotatebox{90}{\textbf{Algorithm\,\,\,}}} & BG-AMP	&	-16.88 (dB) $\mid$ \textbf{09.11 (s)}&	-11.67 (dB) $\mid$ \textbf{08.27 (s)}	& -08.56 (dB) $\mid$ \textbf{06.63 (s)} \\[3 pt] \cline{2-5}
		& GM-AMP ($D = 4$)	&	-17.49 (dB) $\mid$ 19.36 (s)	&	-13.74 (dB) $\mid$ 17.48 (s)	& -10.23 (dB) $\mid$ 15.98 (s) \\[3 pt] \cline{2-5}
		& DCS-BG-AMP	&	-19.84 (dB) $\mid$ 10.20 (s)	&	-14.33 (dB) $\mid$ 08.39 (s)	& -11.40 (dB) $\mid$ 06.71 (s) \\[3 pt] \cline{2-5}
		& DCS-GM-AMP ($D = 4$)	&	\textbf{-21.33 (dB)} $\mid$ 20.34 (s)	&	\textbf{-16.78 (dB)} $\mid$ 18.63 (s)	& \textbf{-12.49 (dB)} $\mid$ 10.13 (s) \\[3 pt] \cline{1-5}
	\end{tabular}
	}
	\caption{Performance on audio CS dataset (TNMSE (dB) $\mid$ Runtime (s)) of two temporally independent algorithms, BG-AMP and GM-AMP, and two temporally structured algorithms, DCS-BG-AMP and DCS-GM-AMP.}
	\label{tab:audio_cs}
\end{table*}

\subsection{Frequency Estimation}
In a final experiment, we compared the performance of DCS-AMP against techniques designed to solve the problem of subspace identification and tracking from partial observations (SITPO) \cite{BNR2010,CEC2012}, which bears similarities to the dynamic CS problem.  In subspace identification, the goal is to learn the low-dimensional subspace occupied by multi-timestep data measured in a high ambient dimension, while in subspace tracking, the goal is to track that subspace as it evolves over time.  In the partial observation setting, the high-dimensional observations are sub-sampled using a mask that varies with time.  The dynamic CS problem can be viewed as a special case of SITPO, wherein the time-$t$ subspace is spanned by a subset of the columns of an a priori known matrix $\vect{A}$.  One problem that lies in the intersection of SITPO and dynamic CS is frequency tracking from partial time-domain observations.

For comparison purposes, we replicated the ``direction of arrival analysis'' experiment described in \cite{CEC2012} where the observations at time $t$ take the form
\begin{equation}
	\vect{y} = \vect{\Phi} \vect{V} \vect{a} + \vect{e}, \quad t = 1,2,\ldots,T
	\label{eq:doa_sensor_measurements}
\end{equation}
where $\vect{\Phi} \in \{0,1\}^{M \times N}$ is a selection matrix with non-zero column indices $\mathcal{Q}^{(t)} \subset \{1,\ldots,N\}$, $\vect{V} \in \mathbb{C}^{N \times K}$ is a Vandermonde matrix of sampled complex sinusoids, i.e.,
\begin{equation}
	\vect{V} \triangleq [\vec{v}(\omega_1^{(t)}),\ldots, \vec{v}(\omega_K^{(t)})],
	\label{eq:vandermonde}
\end{equation}
with $\vec{v}(\omega_k^{(t)}) \triangleq [1, e^{j 2 \pi \omega_k^{(t)}},\ldots, e^{j 2 \pi \omega_k^{(t)} (N-1)}]^{\textsf{T}}$ and $\omega_k^{(t)} \in [0,1)$.  $\vect{a} \in \mathbb{R}^K$ is a vector of instantaneous amplitudes, and $\vect{e} \in \mathbb{R}^N$ is additive noise with i.i.d. $\mathcal{N}(0,\sigma_e^2)$ elements.\footnote{Code for replicating the experiment provided by the authors of \cite{CEC2012}.  \textr{Unless otherwise noted,} specific choices regarding $\{\omega_k^{(t)}\}$ and $\{\vect{a}\}$ were made by the authors of \cite{CEC2012} \textr{in a deterministic fashion}, and can be found in the code. \label{sitpo_foot}}  Here, $\{\vect{\Phi}\}_{t=1}^T$ is known, while $\{\vect{\omega}\}_{t=1}^T$ and $\{\vect{a}\}_{t=1}^T$ are unknown, and our goal is to estimate them.  To assess performance, we report TNMSE in the estimation of the ``complete'' signal $\{\vect{V} \vect{a}\}_{t=1}^T$.

We compared DCS-AMP's performance against two online algorithms designed to solve the SITPO problem: GROUSE \cite{BNR2010} and PETRELS \cite{CEC2012}.  Both GROUSE and PETRELS return time-varying subspace estimates, which were passed to an ESPRIT algorithm to generate time-varying frequency estimates (as in \cite{CEC2012}).  Finally, time-varying amplitude estimates were computed using least-squares.  For DCS-AMP, we constructed $\vect{A}$ using a $2\times$ column-oversampled DFT matrix, keeping only those rows indexed by $\mathcal{Q}^{(t)}$.  DCS-AMP was run in filtering mode for fair comparison with the ``online'' operation of GROUSE and PETRELS, with $I = 7$ inner AMP iterations.

The results of performing the experiment for three different problem configurations are presented in \tabref{freq_estim}, with performance averaged over $100$ independent realizations.  All three algorithms were given the true value of $K$.  In the first problem setup considered, we see that GROUSE operates the fastest, although its TNMSE performance is noticeably inferior to that of both PETRELS and DCS-AMP, which provide similar TNMSE performance and complexity.  In the second problem setup, we reduce the number of measurements, $M$, from $30$ to $10$, leaving all other settings fixed.  In this regime, both GROUSE and PETRELS are unable to accurately estimate $\{\omega_k^{(t)}\}$, and consequently fail to accurately recover $\vect{V} \vect{a}$, in contrast to DCS-AMP.  In the third problem setup, we increased the problem dimensions from the first problem setup by a factor of $4$ to understand how the complexity of each approach scales with problem size.  \textr{In order to increase the number of ``active'' frequencies from $K = 5$ to $K = 20$, $15$ additional frequencies and amplitudes were added uniformly at random to the $5$ deterministic trajectories of the preceding experiments.}  Interestingly, DCS-AMP, which was the slowest at smaller problem dimensions, becomes the fastest (and most accurate) in the higher-dimensional setting, scaling much better than either GROUSE or PETRELS.
\begin{table*}
	\centering
	\ifthenelse{\boolean{ONE_COLUMN}}
	{
	\begin{tabular}{| c | c | c | c | c |}
		\hline
		\multicolumn{2}{|c|}{} & \multicolumn{3}{c|}{\textbf{Problem Setup}} \\ \cline{3-5}
		\multicolumn{2}{|c|}{} & $N = 256$, $M = 30$, $K = 5$	&	$N = 256$, $M = 10$, $K = 5$	&	$N = 1024$, $M = 120$, $K = 20$ \\
		\hline
		\multirow{3}{*}{\rotatebox{90}{\textbf{Algorithm\,\,}}} & GROUSE	&	-4.52 (dB) $\mid$ \textbf{6.78 (s)}&	2.02 (dB) $\mid$ \textbf{6.68 (s)}	& -4.51 (dB) $\mid$ 173.89 (s) \\ \cline{2-5}
		& PETRELS	&	\textbf{-15.62} (dB) $\mid$ 29.51 (s)	&	0.50 (dB) $\mid$ 14.93 (s)	& -7.98 (dB) $\mid$ 381.10 (s) \\ \cline{2-5}
		& DCS-AMP	&	-15.46 (dB) $\mid$ 34.49 (s)	&	\textbf{-10.85 (dB)} $\mid$ 28.42 (s)	& \textbf{-12.79 (dB)} $\mid$ \textbf{138.07 (s)} \\ \cline{1-5}
	\end{tabular}
	}
	{
	\begin{tabular}{| >{\centering\arraybackslash}m{0.25cm} | >{\centering\arraybackslash}m{1.5cm} | >{\centering\arraybackslash}m{3.75cm} | >{\centering\arraybackslash}m{3.75cm} | >{\centering\arraybackslash}m{3.75cm} |}
		\hline
		\multicolumn{2}{|c|}{} & \multicolumn{3}{c|}{\textbf{Problem Setup}} \\ \cline{3-5}
		\multicolumn{2}{|c|}{} & $N = 256$, $M = 30$, $K = 5$	&	$N = 256$, $M = 10$, $K = 5$	&	$N = 1024$, $M = 120$, $K = 20$ \\[3 pt]
		\hline
		\multirow{3}{*}{\rotatebox{90}{\textbf{Algorithm}}} & GROUSE	&	-4.52 (dB) $\mid$ \textbf{6.78 (s)}&	2.02 (dB) $\mid$ \textbf{6.68 (s)}	& -4.51 (dB) $\mid$ 173.89 (s) \\[3 pt] \cline{2-5}
		& PETRELS	&	\textbf{-15.62} (dB) $\mid$ 29.51 (s)	&	0.50 (dB) $\mid$ 14.93 (s)	& -7.98 (dB) $\mid$ 381.10 (s) \\[3 pt] \cline{2-5}
		& DCS-AMP	&	-15.46 (dB) $\mid$ 34.49 (s)	&	\textbf{-10.85 (dB)} $\mid$ 28.42 (s)	& \textbf{-12.79 (dB)} $\mid$ \textbf{138.07 (s)} \\[3 pt] 
		\hline
	\end{tabular}
	}
	\caption{Average performance on synthetic frequency estimation experiment (TNMSE (dB) $\mid$ Runtime (s)) of GROUSE, PETRELS, and DCS-AMP.  In all cases, $T = 4000$, $\sigma_e^2 = 10^{-6}$.}
	\label{tab:freq_estim}
\end{table*}

\section{Conclusion}
\label{sec:conclusion}
In this work we proposed DCS-AMP, a novel approach to dynamic CS.  Our technique merges ideas from the fields of belief propagation and switched linear dynamical systems, together with a computationally efficient inference method known as AMP.  Moreover, we proposed an EM approach that learns all model parameters automatically from the data.  In numerical experiments on synthetic data, DCS-AMP performed within $3$ dB of the support-aware Kalman smoother bound across the sparsity-undersampling plane.  Repeating the dynamic MRI experiment from \cite{LV2009}, DCS-AMP slightly outperformed Modified-CS in MSE, but required less than $10$ seconds to run, in comparison to more than $7$ hours for Modified-CS.  For the compressive sensing of audio, we demonstrated significant gains from the exploitation of temporal structure and Gaussian-mixture learning of the signal prior.  Lastly, we found that DCS-AMP can outperform recent approaches to Subspace Identification and Tracking from Partial Observations (SITPO) when the underlying problem can be well-represented through a dynamic CS model.

\appendices
\section{The Basics of Belief Propagation and AMP}
\label{app:amp_primer}
\textr{In this appendix, we provide a brief primer on belief propagation and the Bayesian approximate message passing (AMP) algorithmic framework proposed by Donoho, Maleki, and Montanari \cite{DMM2010}.  In what follows, we consider the task of estimating a signal vector $\vec{x} \in \mathbb{C}^N$ from linearly compressed and AWGN-corrupted measurements:
\begin{equation}
	\vec{y} = \vec{Ax} + \vec{e} \in \mathbb{C}^M.
	\label{eq:smv_linear_model}
\end{equation}}

\begin{figure}
	\begin{center}
		\ifthenelse{\boolean{ONE_COLUMN}}
		{\scalebox{0.95}{\includegraphics*[1.2in,9.2in][3.3in,10.70in]{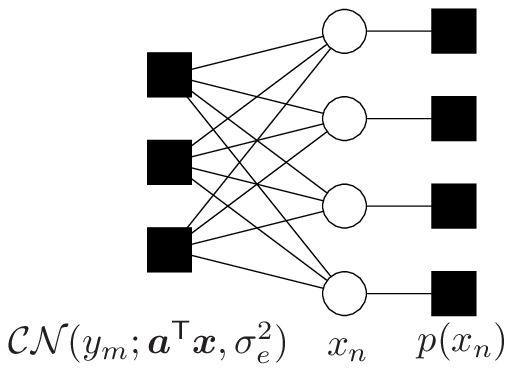}}}
		{\scalebox{0.79}{\includegraphics*[1.2in,9.2in][3.3in,10.70in]{smv_factor_graph1}}}
		\caption{\textr{The factor graph representation of the decomposition of \eqref{smv_post_decomp}.}}
		\label{fig:smv_factor_graph}
	\end{center}
\end{figure}

\textr{AMP can be derived from the perspective of loopy belief propagation (LBP) \cite{KFL2001}, a Bayesian inference strategy that is based on a factorization of the signal posterior pdf, $p(\vec{x}|\vec{y})$, into a product of simpler pdfs that, together, reveal the probabilistic structure in the problem.  Concretely, if the signal coefficients, $\vec{x}$, and noise samples, $\vec{w}$, in \eqref{smv_linear_model} are jointly independent such that $p(\vec{x}) = \prod_{n=1}^N p(x_n)$ and $p(\vec{y} | \vec{x}) = \prod_{m=1}^M \mathcal{CN}(y_m; \vec{a}_m^{\textsf{T}} \vec{x}, \sigma_e^2)$, then the posterior pdf factors as 
\begin{equation}
	p(\vec{x} | \vec{y}) \propto \prod_{m=1}^M \mathcal{CN}(y_m; \vec{a}_m^{\textsf{T}} \vec{x}, \sigma_e^2) \prod_{n=1}^N p(x_n),
	\label{eq:smv_post_decomp}
\end{equation}
yielding the factor graph in \figref{smv_factor_graph}.}

\textr{In belief propagation \cite{P1988}, messages representing beliefs about the unknown variables are exchanged amongst the nodes of the factor graph until convergence to a stable fixed point occurs.  The set of beliefs passed into a given variable node are then used to infer statistical properties of the associated random variable, e.g., the posterior mode, or a complete posterior distribution.  The sum-product algorithm \cite{KFL2001} is perhaps the most well-known approach to belief propagation, wherein the messages take the form of probability distributions, and exact posteriors are guaranteed whenever the graph does not have cycles (``loops''). For graphs with cycles, exact inference is known to be NP-hard, and so LBP is not guaranteed to produce correct posteriors.  Still, it has shown state-of-the-art performance on a wide array of challenging inference problems, as noted in \secref{algorithm:implementation}.}

\textr{The conventional wisdom surrounding LBP says that accurate inference is possible only when the factor graph is locally tree-like, i.e., the girth of any cycle is relatively large.  With \eqref{smv_linear_model}, this would require that $\vec{A}$ is an appropriately constructed sparse matrix, which precludes some of the most interesting CS problems.  In a remarkable departure from convention, Donoho, Maleki, Montanari, and Bayati demonstrated that LBP-based compressive sensing is not only feasible \cite{DMM2009,DMM2010} for dense $\vec{A}$ matrices, but provably accurate \cite{BM2011}.  In particular, they established that, in the large-system limit (i.e., as $M,N \to \infty$ with $M/N$ fixed) and under i.i.d. sub-Gaussian $\vec{A}$, the iterations of AMP are governed by a state-evolution whose fixed point---when unique---yields the true posterior means.  Beyond its theoretical significance, AMP is important for its computational properties as well.  As demonstrated in the original AMP work \cite{DMM2009}, not only can LBP solve the compressive sensing problem \eqref{smv_linear_model}, but it can do so much faster, and more accurately, than other state-of-the-art methods, whether optimization-based, greedy, or Bayesian. To accomplish this feat, \cite{DMM2009,DMM2010} proposed a specific set of approximations that become accurate in the limit of large, dense $\vec{A}$ matrices, yielding algorithms that give accurate results using only $\approx2MN$ flops-per-iteration, and relatively few iterations (e.g., tens).}

\textr{The specific implementation of any AMP algorithm will depend on the particular choices of likelihood and prior, but ultimately amounts to an iterative, scalar soft-thresholding procedure with a carefully chosen adaptive thresholding strategy.  Deriving the appropriate thresholding functions for a particular signal model can be accomplished by computing scalar sum-product, or max-sum, updates of a simple form (see, e.g., \cite[Table 1]{R2010b}).}

\bibliographystyle{ieeetr}
\bibliography{time_evolve_sparse.bib}

\begin{thebibliography}{10}

\bibitem{ZPS2010}
J.~Ziniel, L.~C. Potter, and P.~Schniter, ``Tracking and smoothing of
  time-varying sparse signals via approximate belief propagation,'' in {\em
  Asilomar Conf. on Signals, Systems and Computers 2010}, (Pacific Grove, CA),
  Nov. 2010.

\bibitem{V2008}
N.~Vaswani, ``Kalman filtered compressed sensing,'' in {\em IEEE Int'l Conf. on
  Image Processing (ICIP) 2008}, pp.~893 --896, 12-15 2008.

\bibitem{ARBV2010}
M.~Salman~Asif, D.~Reddy, P.~Boufounos, and A.~Veeraraghavan, ``Streaming
  compressive sensing for high-speed periodic videos,'' in {\em Int'l Conf. on
  Image Processing (ICIP) 2010}, (Hong Kong), Sept. 2010.

\bibitem{LP2007}
W.~Li and J.~C. Preisig, ``Estimation of rapidly time-varying sparse
  channels,'' {\em IEEE J. Oceanic Engr.}, vol.~32, pp.~927--939, Oct. 2007.

\bibitem{AGG2009}
D.~Angelosante, G.~B. Giannakis, and E.~Grossi, ``Compressed sensing of
  time-varying signals,'' in {\em Int'l Conf. on Digital Signal Processing
  2009}, pp.~1--8, July 2009.

\bibitem{V2010b}
N.~Vaswani, ``{LS-CS-residual} {(LS-CS)}: {C}ompressive sensing on least
  squares residual,'' {\em IEEE Trans. Signal Process.}, vol.~58, no.~8,
  pp.~4108--4120, 2010.

\bibitem{DV2010}
S.~Das and N.~Vaswani, ``Particle filtered modified compressive sensing
  {(PaFiMoCS)} for tracking signal sequences,'' in {\em Asilomar Conf. on
  Signals, Systems and Computers 2010}, pp.~354--358, Nov. 2010.

\bibitem{LV2012}
W.~Lu and N.~Vaswani, ``Regularized modified bpdn for noisy sparse
  reconstruction with partial erroneous support and signal value knowledge,''
  {\em Signal Processing, IEEE Transactions on}, vol.~60, no.~1, pp.~182--196,
  2012.

\bibitem{VL2010}
N.~Vaswani and W.~Lu, ``Modified-{CS}: {M}odifying compressive sensing for
  problems with partially known support,'' {\em IEEE Trans. Signal Process.},
  vol.~58, pp.~4595--4607, Sept. 2010.

\bibitem{ARG2009}
D.~Angelosante, S.~Roumeliotis, and G.~Giannakis, ``Lasso-{K}alman smoother for
  tracking sparse signals,'' in {\em Asilomar Conf. on Signals, Systems and
  Computers 2009}, (Pacific Grove, CA), pp.~181 -- 185, Nov. 2009.

\bibitem{DSM2011}
W.~Dai, D.~Sejdinovi\'c, and O.~Milenkovic, ``Gaussian dynamic compressive
  sensing,'' in {\em Int'l Conf. on Sampling Theory and Appl. (SampTA)},
  (Singapore), May 2011.

\bibitem{SAP2010}
D.~Sejdinovi\'c, C.~Andrieu, and R.~Piechocki, ``Bayesian sequential compressed
  sensing in sparse dynamical systems,'' in {\em 48th Allerton Conf. Comm.,
  Control, \& Comp.}, (Urbana, IL), pp.~1730--1736, Nov. 2010.

\bibitem{STR2011}
B.~Shahrasbi, A.~Talari, and N.~Rahnavard, ``{TC-CSBP}: {C}ompressive sensing
  for time-correlated data based on belief propagation,'' in {\em Conf. on
  Inform. Sci. and Syst. (CISS)}, (Baltimore, MD), pp.~1--6, Mar. 2011.

\bibitem{RC2004}
C.~P. Robert and G.~Casella, {\em Monte {C}arlo Statistical Methods}.
\newblock Springer, 2nd~ed., 2004.

\bibitem{FM1998}
B.~J. Frey and D.~J.~C. MacKay, ``A revolution: {B}elief propagation in graphs
  with cycles,'' {\em Neural Information Processing Systems (NIPS)}, vol.~10,
  pp.~479--485, 1998.

\bibitem{BSB2010}
D.~Baron, S.~Sarvotham, and R.~G. Baraniuk, ``Bayesian compressive sensing via
  belief propagation,'' {\em IEEE Trans. Signal Process.}, vol.~58, pp.~269
  --280, Jan. 2010.

\bibitem{DMM2009}
D.~L. Donoho, A.~Maleki, and A.~Montanari, ``Message passing algorithms for
  compressed sensing,'' in {\em Proceedings of the National Academy of
  Sciences}, vol.~106, pp.~18914--18919, Nov. 2009.

\bibitem{S2010a}
P.~Schniter, ``Turbo reconstruction of structured sparse signals,'' in {\em
  Conf. on Information Sciences and Systems (CISS)}, pp.~1 --6, Mar. 2010.

\bibitem{BM2011}
M.~Bayati and A.~Montanari, ``The dynamics of message passing on dense graphs,
  with applications to compressed sensing,'' {\em IEEE Trans. Inform. Theory},
  vol.~57, pp.~764--785, Feb. 2011.

\bibitem{SS2012}
S.~Som and P.~Schniter, ``Compressive imaging using approximate message passing
  and a {M}arkov-tree prior,'' {\em IEEE Trans. Signal Process.}, vol.~60,
  pp.~3439--3448, Jul. 2012.

\bibitem{S2011}
P.~Schniter, ``A message-passing receiver for {BICM-OFDM} over unknown
  clustered-sparse channels,'' {\em IEEE J. Select. Topics in Signal Process.},
  vol.~5, pp.~1462--1474, Dec. 2011.

\bibitem{DLR1977}
A.~P. Dempster, N.~M. Laird, and D.~B. Rubin, ``Maximum likelihood from
  incomplete data via the {EM} algorithm,'' {\em J. R. Statist. Soc. B},
  vol.~39, pp.~1--38, 1977.

\bibitem{KFL2001}
F.~R. Kschischang, B.~J. Frey, and H.~A. Loeliger, ``Factor graphs and the
  sum-product algorithm,'' {\em IEEE Trans. Inform. Theory}, vol.~47,
  pp.~498--519, Feb. 2001.

\bibitem{P1988}
J.~Pearl, {\em Probabilistic Reasoning in Intelligent Systems}.
\newblock San Mateo, CA: Morgan Kaufman, 1988.

\bibitem{WF2001}
Y.~Weiss and W.~T. Freeman, ``Correctness of belief propagation in {G}aussian
  graphical models of arbitrary topology,'' {\em Neural Computation}, vol.~13,
  pp.~2173--2200, Oct. 2001.

\bibitem{TJ2002}
S.~C. Tatikonda and M.~I. Jordan, {\em Loopy belief propagation and {G}ibbs
  measures}, pp.~493--500.
\newblock Proc. 18th Conf. Uncertainty in Artificial Intelligence (UAI), San
  Mateo, CA: Morgan Kaufmann, 2002.

\bibitem{H2004}
T.~Heskes, ``On the uniqueness of belief propagation fixed points,'' {\em
  Neural Comput.}, vol.~16, no.~11, pp.~2379--2413, 2004.

\bibitem{IFW2005}
A.~T. Ihler, J.~W. Fisher~III, and A.~S. Willsky, ``Loopy belief propagation:
  {C}onvergence and effects of message errors,'' {\em J. Mach. Learn. Res.},
  vol.~6, pp.~905--936, 2005.

\bibitem{MMC1998}
R.~J. McEliece, D.~J.~C. MacKay, and J.~Cheng, ``Turbo decoding as an instance
  of {P}earl's belief propagation algorithm,'' {\em IEEE J. Select. Areas
  Comm.}, vol.~16, pp.~140--152, Feb. 1998.

\bibitem{FPC2000}
W.~T. Freeman, E.~C. Pasztor, and O.~T. Carmichael, ``Learning low-level
  vision,'' {\em Int'l. J. Comp. Vision}, vol.~40, pp.~25--47, Oct. 2000.

\bibitem{DMM2010}
D.~L. Donoho, A.~Maleki, and A.~Montanari, ``Message passing algorithms for
  compressed sensing: I. motivation and construction,'' in {\em Proc. of
  Information Theory Workshop}, Jan. 2010.

\bibitem{ZS2013}
J.~Ziniel and P.~Schniter, ``Efficient high-dimensional inference in the
  multiple measurement vector problem,'' {\em IEEE Trans. Signal Process.},
  vol.~61, pp.~340--354, Jan. 2013.

\bibitem{VS2012}
J.~P. Vila and P.~Schniter, ``Expectation-{M}aximization {G}aussian-mixture
  approximate message passing,'' in {\em Proc. Conf. on Information Sciences
  and Systems}, (Princeton, NJ), Mar. 2012.

\bibitem{EMK2006}
G.~Elidan, I.~McGraw, and D.~Koller, ``Residual belief propagation: {I}nformed
  scheduling for asynchronous message passing,'' in {\em Proc. 22nd Conf.
  Uncertainty Artificial Intelligence (UAI)}, 2006.

\bibitem{R2011}
S.~Rangan, ``Generalized approximate message passing for estimation with random
  linear mixing,'' in {\em Proc. IEEE Int'l Symp. Inform. Theory}, (St.
  Petersburg, Russia), pp.~2168--2172, Aug. 2011.
\newblock (See also arXiv: 1010.5141).

\bibitem{R2010b}
S.~Rangan, ``Generalized approximate message passing for estimation with random
  linear mixing.'' arXiv:1010.5141v1 [cs.IT], 2010.

\bibitem{SS2011}
S.~Som and P.~Schniter, ``Approximate message passing for recovery of sparse
  signals with {M}arkov-random-field support structure,'' in {\em Int'l Conf.
  Mach. Learn.}, (Bellevue, Wash.), Jul. 2011.

\bibitem{ZRS2012}
J.~Ziniel, S.~Rangan, and P.~Schniter, ``A generalized framework for learning
  and recovery of structured sparse signals,'' in {\em Proc. Stat. Signal
  Process. Wkshp}, (Ann Arbor, MI), Aug. 2012.

\bibitem{ZH2005}
O.~Zoeter and T.~Heskes, ``Change point problems in linear dynamical systems,''
  {\em J. Mach. Learn. Res.}, vol.~6, pp.~1999--2026, Dec. 2005.

\bibitem{AS1972}
D.~L. Alspach and H.~W. Sorenson, ``Nonlinear {B}ayesian estimation using
  {G}aussian sum approximations,'' {\em IEEE Trans. Auto. Control}, vol.~17,
  pp.~439--448, Aug. 1972.

\bibitem{BC2010}
D.~Barber and A.~T. Cemgil, ``Graphical models for time series,'' {\em IEEE
  Signal Process. Mag.}, vol.~27, pp.~18--28, Nov. 2010.

\bibitem{Z2013}
J.~Ziniel, {\em Message Passing Approaches to Compressive Inference Under
  Structured Signal Priors}.
\newblock PhD thesis, The Ohio State University, 2013.
\newblock In preparation.

\bibitem{M1996}
T.~K. Moon, ``The expectation-maximization algorithm,'' {\em IEEE Signal
  Process. Mag.}, vol.~13, pp.~47--60, Nov. 1996.

\bibitem{B2006}
C.~M. Bishop, {\em Pattern Recognition and Machine Learning}.
\newblock New York: Springer-Verlag, 2006.

\bibitem{DT2009}
D.~L. Donoho and J.~Tanner, ``Observed universality of phase transitions in
  high-dimensional geometry, with implications for modern data analysis and
  signal processing,'' {\em Phil. Trans. R. Soc. A}, vol.~367, pp.~4273--4293,
  2009.

\bibitem{E2006}
R.~L. Eubank, {\em A {K}alman Filter Primer}.
\newblock Boca Raton, FL: Chapman \& Hall/CRC, 2006.

\bibitem{LDHKPK2007}
H.~A. Loeliger, J.~Dauwels, J.~Hu, S.~Korl, L.~Ping, and F.~R. Kschischang,
  ``The factor graph approach to model-based signal processing,'' {\em Proc. of
  the IEEE}, vol.~95, no.~6, pp.~1295--1322, 2007.

\bibitem{CDS1998}
S.~S. Chen, D.~L. Donoho, and M.~A. Saunders, ``Atomic decomposition by basis
  pursuit,'' {\em SIAM J. Scientific Comp.}, vol.~20, no.~1, pp.~33--61, 1998.

\bibitem{BF2008}
E.~van~den Berg and M.~Friedlander, ``Probing the {P}areto frontier for basis
  pursuit solutions,'' {\em SIAM J. Scientific Comp.}, vol.~31, pp.~890--912,
  Nov. 2008.

\bibitem{ZR2011}
Z.~Zhang and B.~D. Rao, ``Sparse signal recovery with temporally correlated
  source vectors using {S}parse {B}ayesian {L}earning,'' {\em IEEE J. Selected
  Topics Signal Process.}, vol.~5, pp.~912--926, Sept. 2011.

\bibitem{LV2009}
W.~Lu and N.~Vaswani, ``Modified compressive sensing for real-time dynamic {MR}
  imaging,'' in {\em Image Processing (ICIP), 2009 IEEE Int'l Conf. on},
  pp.~3045 --3048, Nov. 2009.

\bibitem{BNR2010}
L.~Balzano, R.~Nowak, and B.~Recht, ``Online identification and tracking of
  subspaces from highly incomplete information,'' in {\em Allerton Conf. on
  Comm., Control, and Comp.}, pp.~704 --711, Oct. 2010.

\bibitem{CEC2012}
Y.~Chi, Y.~Eldar, and R.~Calderbank, ``{PETRELS}: {S}ubspace estimation and
  tracking from partial observations,'' in {\em Int'l Conf. Acoustics, Speech,
  \& Signal Process. (ICASSP)}, (Kyoto, Japan), Mar. 2012.

\end{thebibliography}

\end{document}